\documentclass[10pt,journal,epsfig]{IEEEtran}
\usepackage{graphicx}
\usepackage{subfigure,color}
\usepackage{epstopdf}
\usepackage{amssymb}
\usepackage{cite,comment}
\usepackage{amsmath}
\usepackage{bm,comment,setspace}
\usepackage{algorithm}
\usepackage{algpseudocode}
\makeatletter

\newcommand{\Rmnum}[1]{\expandafter\@slowromancap\romannumeral #1@}
\newcommand{\mv}[1]{\mbox{\boldmath{$ #1 $}}}

\makeatother

\newtheorem{proposition}{\underline{Proposition}}
\newtheorem{remark}{\underline{Remark}}

\newtheorem{lemma}{\underline{Lemma}}

\begin{document}
\title{Cooperative Downlink Interference Transmission and Cancellation for Cellular-Connected UAV: A Divide-and-Conquer Approach}
\author{Weidong Mei and Rui Zhang, \IEEEmembership{Fellow, IEEE}
\thanks{\scriptsize{This work will be presented in part at the IEEE Global Communications Conference, December 9-13, 2018, Waikoloa, Hawaii, USA\cite{cooperative2019mei}.}}
\thanks{\scriptsize{W. Mei is with the NUS Graduate School for Integrative Sciences and Engineering, National University of Singapore, Singapore 119077, and also with the Department of Electrical and Computer Engineering, National University of Singapore, Singapore 117583 (e-mail: wmei@u.nus.edu).}}
\thanks{\scriptsize{R. Zhang is with the Department of Electrical and Computer Engineering, National University of Singapore, Singapore 117583 (e-mail: elezhang@nus.edu.sg).}}}
\maketitle

\begin{abstract}
The line-of-sight (LoS) dominant air-ground channels have posed critical interference issues in cellular-connected unmanned aerial vehicle (UAV) communications. In this paper, we propose a new base station (BS) cooperative beamforming (CB) technique for the cellular downlink to mitigate the strong interference caused by the co-channel terrestrial transmissions to the UAV. Besides the conventional CB by cooperatively transmitting the UAV's message, the serving BSs of the UAV exploit a novel CB-based {\it \textbf{interference transmission}} scheme to effectively suppress the terrestrial interference to the UAV. Specifically, the co-channel terrestrial users' messages are shared with the UAV's serving BSs and transmitted via CB so as to cancel their resultant interference at the UAV's receiver. To optimally balance between the CB gains for UAV signal enhancement and terrestrial interference cancellation, we formulate a new problem to maximize the UAV's receive signal-to-interference-plus-noise ratio (SINR) by jointly optimizing the power allocations at all of its serving BSs for transmitting the UAV's and co-channel terrestrial users' messages. First, we derive the closed-form optimal solution to this problem in the special case of one serving BS for the UAV and draw useful insights. Then, we propose an algorithm to solve the problem optimally in the general case. As the optimal solution requires centralized implementation with exorbitant message/channel information exchanges among the BSs, we further propose a distributed algorithm that is amenable to practical implementation, based on a new {\it divide-and-conquer} approach, whereby each co-channel BS divides its perceived interference to the UAV into multiple portions, each to be canceled by a different serving BS of the UAV with its best effort. Numerical results show that the proposed centralized and distributed CB schemes with interference transmission and cancellation (ITC) can both significantly improve the UAV's downlink performance as compared to the conventional CB without applying ITC.
\end{abstract}
\begin{IEEEkeywords}
Unmanned aerial vehicle (UAV), cellular-connected UAV, cooperative beamforming, interference transmission and cancellation, distributed algorithm.
\end{IEEEkeywords}

\section{Introduction}
Thanks to their swift deployment and controllable mobility, unmanned aerial vehicles (UAVs) (a.k.a. drones) have been gaining increasing popularity in recent years, not just as a toy-grade equipment for hobbyists, but more importantly as the enabler for a plethora of new applications, such as cargo delivery, surveillance and inspection, aerial photography, among others\cite{zeng2019accessing}. Recent statistics by the Federal Aviation Administration (FAA) show that more than 100,000 people have obtained a remote pilot certificate to fly a drone for commercial and recreational uses as of July 2018. The rapidly evolving and booming UAV market is also enticing wireless communication industry to join the UAV ``gold rush''. On one hand, advances in communication equipment miniaturization have enabled UAVs to serve as communication platforms in the sky (such as quasi-stationary and mobile base stations (BSs)/relays), to provide or enhance the communication services for the terrestrial or even aerial user equipments (UEs) in demand\cite{zeng2016wireless,sekander2018multi,li2018uav,athanasiadou2019lte,mozaffari2018beyond}.
On the other hand, to support the large-scale deployment of UAVs in the future, an appealing solution is by integrating UAVs into the future cellular network (i.e., the fifth generation (5G) and beyond) as new aerial UEs that are able to communicate with the terrestrial BSs. Compared to the existing UAV-ground communications available only within the pilot's visual line-of-sight (LoS) range, cellular-connected UAVs are enabled by the beyond visual and radio LoS (BVRLoS) communications, with significant performance enhancement in terms of reliability, coverage, security and throughput\cite{zeng2019cellular,bor20195g}. In fact, several preliminary field trials have demonstrated that it is feasible to support the basic communication requirements for UAVs with today's fourth generation (4G) or Long Term Evolution (LTE) network\cite{qualcom2017lte,lin2018sky}.

Despite the above advantages, integrating UAVs into future cellular networks faces new challenges. In particular, how to mitigate the severe aerial-ground interference is deemed as a major challenge in practically realizing cellular-connected UAVs. Compared to terrestrial wireless channels that in general suffer from more severe path-loss, shadowing and multi-path fading, the high altitude of UAVs generally leads to LoS-dominant channels with ground BSs. Due to the LoS links, the UAV may cause/suffer more severe uplink/downlink interference to/from a much larger number of BSs than ground UEs, which could significantly degrade the communication performance of UAVs in the downlink as well as that of ground UEs in the uplink. Although various interference mitigation techniques have been studied in the literature among which some were applied to the terrestrial networks (such as inter-cell interference coordination (ICIC)\cite{boudreau2009interference,kosta2013interference,hamza2013survey}, coordinated multi-point (CoMP) transmission\cite{sawahashi2010coordinated,lee2012coordinated,lee2012comp} and non-orthogonal multiple access (NOMA)\cite{saito2013non,islam2017power,ding2017survey}), they may be ineffective or insufficient to deal with the new and more severe interference issue brought by UAVs, owing to their unique LoS-dominant air-ground channels.
As such, new and more sophisticated interference mitigation techniques are needed to achieve efficient spectrum sharing between the existing ground UEs and new aerial UEs in future cellular networks. There have been several recent works \cite{zeng2019cellular,kovacs2017interference,yaj2018interference,geraci2018understanding,amorim2018measured,liu2018multi,cellular2018mei,mei2019uplink} devoted to this new direction. In \cite{zeng2019cellular,kovacs2017interference,yaj2018interference,geraci2018understanding,amorim2018measured}, the authors evaluated the performance of several existing techniques, such as three-dimensional (3D) beamforming, closed-loop power control, and massive multiple-input multiple-output (MIMO) for UAV communications via simulations and/or measurements. In contrast, the authors in \cite{liu2018multi,cellular2018mei,mei2019uplink} proposed new and enhanced interference mitigation techniques for cellular-connected UAV communications. Specifically, the authors in \cite{liu2018multi} first proposed a multi-beam UAV communication scheme for cellular uplink, where a new form of cooperative interference cancellation by exploiting the idle/available terrestrial BSs was applied jointly with UAV's transmit beamforming to mitigate its strong uplink interference to the ground UEs. In \cite{cellular2018mei}, new aerial-ground ICIC designs were proposed for UAV uplink communication to maximize the network throughput by treating the interference as noise. In \cite{mei2019uplink}, a novel cooperative NOMA strategy was proposed to further improve the performance of the ICIC design in \cite{cellular2018mei}, by employing the interference cancellation at cooperative BSs. The work \cite{huang2019cognitive} aimed to mitigate the UAV's uplink interference via joint maneuver and power control, subject to its interference power constraints at the ground BSs. However, none of the above works\cite{liu2018multi,cellular2018mei,mei2019uplink,huang2019cognitive} has addressed how to deal with the interference issue in the cellular downlink communication with UAVs, and their proposed interference mitigation techniques may be not applicable to the cellular downlink, due to the fundamental role change of the UAV from an interference source to an interference victim in cellular networks.

\begin{figure}[!t]
\centering
\includegraphics[width=3.2in]{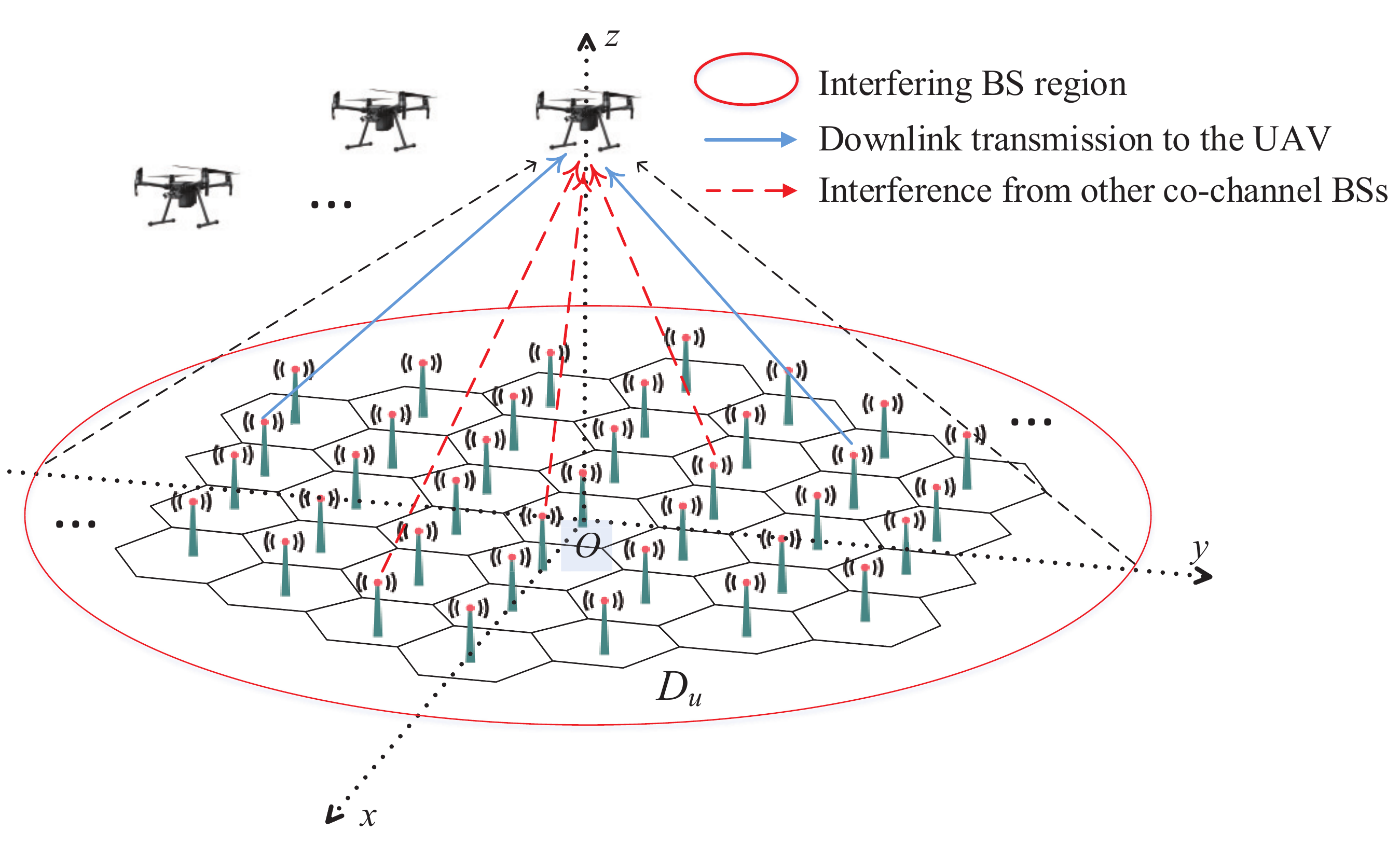}
\DeclareGraphicsExtensions.
\caption{Downlink UAV communications in a cellular network.}\label{down}
\vspace{-12pt}
\end{figure}
Motivated by the above, this paper investigates the downlink interference mitigation solution in a cellular network with co-existing UAVs and ground UEs. As shown in Fig.\,\ref{down}, thanks to the strong air-ground LoS channels, each UAV can be associated with multiple ground BSs even far away from its located cell at the same time, thus yielding a higher macro-diversity gain than ground UEs. However, on the other hand, it also suffers more severe inter-cell interference (ICI) from a larger number of non-associated co-channel BSs in a much wider area (denoted as the interfering BS region $D_u$ in Fig.\,\ref{down}) as compared to ground UEs, which can practically result in unsatisfactorily low signal-to-interference-plus-noise ratio (SINR) at the UAV receiver in the downlink. To mitigate the strong interference to the UAV yet without affecting transmissions to the existing ground UEs, a practical solution is cooperative beamforming (CB), where the available BSs in Fig.\,\ref{down} that are not serving any ground UEs in the UAV's assigned time-frequency resource block (RB) transmit to the UAV cooperatively \cite{mudumbai2009distributed}, so as to enhance its received signal power to overcome the strong co-channel interference. However, due to the frequency reuse of ground UEs, the number of available BSs decreases rapidly with increasing ground UE density while the co-channel interference also increases; as a result, the SINR gain of conventional CB diminishes. Moreover, increasing transmit power of all BSs can only marginally improve the UAV's receive SINR since this increases both the UAV's received signal power and interference power at the same time. Although a more effective CB scheme could engage all BSs to transmit cooperatively to their served UAVs and ground UEs in the same RB simultaneously, it requires large-scale CoMP transmission of all co-channel BSs in the UAV's interfering BS region (see Fig.\,\ref{down})\cite{zhang2010cooperative}, which is difficult to implement in practice due to the high complexity and excessive message/channel information exchange among the large number of BSs involved. In order to improve the UAV downlink SINR with practically affordable complexity, this paper proposes a new CB design that is different from the above schemes. Specifically, by leveraging the backhaul links among BSs (e.g., the existing X2 interface in LTE\cite{dahlman20134g,3GPP36423}), the messages for the terrestrial users in the same RB as the UAV are first shared by their BSs to the serving BSs of the UAV and then transmitted by them via CB so as to cancel the co-channel interference at the UAV receiver, along with their cooperatively transmitted UAV's message. By this means, the existing terrestrial transmissions are unaffected while the UAV receiver attains the CB gains for both signal power enhancement and terrestrial interference suppression, thus leading to a significant SINR improvement. We refer to this new scheme as CB with interference transmission and cancellation (ITC).

Interestingly, we show that there exists a fundamental trade-off between the CB gains for UAV signal enhancement and terrestrial ITC. To optimally reconcile this trade-off, we formulate a new problem to maximize the UAV's receive SINR by jointly optimizing the power allocations at all of its serving BSs for transmitting its own as well as the co-channel terrestrial UEs' messages, subject to the per-BS power constraints. First, we derive the closed-form optimal solution to this problem in the special case of one single serving BS of the UAV and draw useful insights to the optimal ITC design. Then, we propose the optimal algorithm to solve the UAV SINR maximization problem in the general case by exploiting its hidden convexity. As the optimal ITC solution needs to be implemented by the UAV's serving BSs in a centralized manner with exorbitant message/channel information shared by all the co-channel BSs, we further propose a distributed algorithm that is amenable to practical implementation, based on a new {\it divide-and-conquer} approach. Specifically, each of the co-channel BSs serving terrestrial users splits its perceived interference to the UAV into multiple portions, each to be canceled by a nearby serving BS of the UAV via ITC with its best effort. The interference splitting ratios can be further updated by each co-channel BS independently to improve the ITC performance based on the feedback from the UAV's serving BSs. Finally, based on the channel models recommended by the 3rd Generation Partnership Project (3GPP)\cite{3GPP36777}, simulation results are provided to show the significant performance gains of the proposed centralized and distributed CB schemes with ITC over the conventional CB without applying ITC.

The rest of this paper is organized as follows. Section \Rmnum{2} presents the system model as well as the conventional CB without ITC and the proposed CB with ITC. Section \Rmnum{3} presents the problem formulation for optimally designing the proposed CB scheme with ITC. Section \Rmnum{4} presents the closed-form optimal solution to this problem in some special cases to draw useful insights, as well as an algorithm that solves our problem optimally in the general case. Section \Rmnum{5} presents a distributed CB scheme for practical implementation based on the divide-and-conquer approach. Section \Rmnum{6} presents the simulation results to show the performance of the proposed CB schemes as compared to benchmark schemes. Finally, Section \Rmnum{7} concludes this paper and discusses future work.

The following main notations are used in this paper. Bold symbols in capital letter and small letter denote matrices and vectors, respectively. The transpose and conjugate transpose of a matrix are denoted as ${(\cdot)}^{T}$ and ${(\cdot)}^{H}$, respectively. ${\mathbb{R}}^n$ (${\mathbb{C}}^n$) denotes the set of real (complex) vectors of length $n$. For complex/real number $s$, $\lvert \cdot \rvert$ denotes the absolute value. For a vector ${\mv a} \in {\mathbb{R}}^n$, ${\mv a} \succeq \mathbf 0$ means that ${\mv a}$ is element-wise nonnegative. ${\mathbb E}[\cdot]$ denotes the expected value of random variables. $x \sim \mathcal{CN}(\mu,\sigma^2)$ means that $x$ is a circularly symmetric complex Gaussian (CSCG) random variable with mean $\mu$ and variance $\sigma^2$. $\lVert \cdot \rVert$ represents the vector Euclidean norm. ${\mv I}_n$ denotes an $n \times n$ identity matrix. $\emptyset$ denotes an empty set. $\lvert A \rvert$ denotes the cardinality of a set $A$. For two sets $A$ and $B$, $A \cap B$ denotes the intersection of $A$ and $B$, $A \cup B$ denotes the union of $A$ and $B$, and $A \backslash B$ denotes the set of elements that belong to $A$ but are not in $B$.

\section{System Model}
As shown in Fig.\,\ref{down}, we consider the downlink communication in a given subregion of the cellular network, where the terrestrial BSs serve multiple UAV UEs and a set of ground UEs. For the purpose of exposition, it is assumed that each UAV is equipped with a single antenna, while each BS employs an antenna array with fixed directional gain pattern\cite{3GPP36777}. We assume that these UAVs are assigned with orthogonal RBs for their downlink communications, each with one or more serving BSs in this subregion, by exploiting the macro-diversity with ground BSs\footnote{To realize the assumed orthogonal RB allocations over UAV UEs, one practically efficient method is that each UAV broadcasts the index of the RB assigned to it by its serving BS to all other BSs in $D_u$ over its uplink control channel, by leveraging its reliable LoS-dominant links with the ground BSs, so that they will not assign this RB to their served UAV UEs.}. Due to frequency reuse in the cellular network, the assigned RB to each UAV is likely to be already used by some BSs in this subregion for serving their respective terrestrial UEs\footnote{This makes it practically unlikely to find an exclusive RB for each UAV to completely avoid the terrestrial interference, especially when the terrestrial traffic load is high in the considered region.}, thus causing strong interference to the UAV over their LoS-dominant channels with it. Without loss of generality, in the sequel of this paper, we consider one particular UAV above the center of $D_u$ in Fig.\,\ref{down} to investigate our proposed scheme, which can be similarly applied to other UAVs over the orthogonal RBs allocated to them. Centered at the UAV's horizontal location projected on the ground, we consider there are in total $J$ BSs located in the interfering BS region $D_u$ of the UAV, as shown in Fig.\,\ref{down}. For BSs outside $D_u$, we assume that their interference is attenuated to the level below the receiver noise at the UAV and thus can be ignored.

Next, we describe the considered cellular network before and after the UAV UE is added. In particular, the UAV's receive SINRs under the schemes without applying CB or with the conventional CB are first derived. Then, we introduce the proposed new CB scheme with ITC and derive its achievable SINR.

\subsection{Cellular Network with Terrestrial UEs Only}\label{criterion}
Denote by $K$ the total number of terrestrial UEs in $D_u$ that are communicating with their associated BSs in the downlink in the same RB assigned to the UAV, with $K \le J$ (because each BS serves at most one UE in each RB). To mitigate their ICI, we assume that the following rule is used when assigning this RB to the $K$ terrestrial UEs by their associated BSs\cite{cellular2018mei}. Let ${\cal N}_j(q)$ denote the set of the first $q$-tier neighboring BSs of BS $j$\footnote{In the case of hexagon cell shape, the first $q$-tier neighboring BSs of BS $j \in \cal J$ refer to all BSs in the first $q$ rings around BS $j$.} including itself, $j \in \{1,2,\cdots,J\}$. If the RB is already occupied by a terrestrial UE in ${\cal N}_j(q)$, then BS $j$ will not assign this RB to any new UEs. By this means, BS $j$ avoids causing any interference to the existing UEs in ${\cal N}_j(q)$ in the same RB. Note that when $q$ is sufficiently large, the terrestrial ICI becomes negligible, thanks to the significant path-loss and shadowing of typical terrestrial channels in rich-scattering environment, as well as the highly restricted frequency reuse.

For convenience, we denote ${\cal J}\triangleq\{1,2,\cdots,J\}$ as the set of all the BSs in $D_u$. Define a set ${\cal J}_o \subseteq \cal J$ with $\lvert {\cal J}_o \rvert=K$, where $j \in {\cal J}_o$ if BS $j$ is currently serving a terrestrial UE (thus termed \emph{occupied} BS) in the same RB of the UAV (to be assigned), and thus ${\cal J}^c_o={\cal J}\backslash {\cal J}_o$. For the purpose of exposition, we assume that the transmit powers of all BSs in ${\cal J}$ over the considered RB equal to $P$, if they transmit to their associated UEs.

\subsection{Cellular Network with New UAV Added}\label{new.uav}
Let $f_j$ be the complex-valued baseband equivalent channel coefficient from BS $j, j \in \cal J$, to the UAV of our interest. The UAV can be associated with any unoccupied BS in the set ${\cal J}_o^c$. To avoid degrading the rate performance of the existing terrestrial UEs in the same RB, an unoccupied BS $n \in {\cal J}_o^c$ is allowed to serve the UAV (thus termed {\it available} BS) if and only if there are no occupied BSs in its first $q$-tier neighborhood, i.e.,
\begin{equation}\label{serv}
{\cal N}_n(q) \cap {\cal J}_o = \emptyset.
\end{equation}
Let $\Omega$ be the set of all the available BSs in $D_u$ with $\lvert \Omega \rvert=N$, and $N \le J-K$.

Obviously, among the $N$ available BSs in $\Omega$, the UAV should associate with the one having the largest channel power gain with it, denoted as $i = \arg \mathop {\max}\limits_{n \in \Omega} \lvert f_n \rvert^2$, if only one single BS is allowed to serve its downlink communication. In this case, the received signal at the UAV can be expressed as
\begin{equation}
y_u=\sqrt{P}f_ix_u+\sum\limits_{j \in {\cal J}_o} {\sqrt{P}f_jx_j}+z_u,
\end{equation}
where $x_u$ and $x_j$ denote the transmitted data symbol for the UAV and that for the terrestrial UE served by BS $j$ with ${\mathbb E}[{\lvert{x_u}\rvert}^2] = 1$ and ${\mathbb E}[{\lvert{x_j}\rvert}^2] = 1, j \in {\cal J}_o$, respectively, and $z_u \sim \mathcal{CN}(0,\sigma^2)$ is the UAV receiver noise, with $\sigma^2$ denoting the noise power. As a result, the UAV's receive SINR is expressed as
\begin{equation}\label{sinr1}
\gamma_u = \frac{P\lvert f_i \rvert^2}{\sigma^2 + P\sum\limits_{j \in {\cal J}_o} {\lvert f_j \rvert^2}}.
\end{equation}
As observed from (\ref{sinr1}), due to the strong LoS-dominant channels, $\lvert f_j\rvert^2$'s can be comparable with $\lvert f_i \rvert^2$ and as a result, the total terrestrial interference at the UAV may significantly overwhelm its desired signal if $\sum\nolimits_{j \in {\cal J}_o} {\lvert f_j \rvert^2} \gg {\lvert f_i \rvert^2}$. In this case, the UAV's SINR and thus achievable rate could be extremely low.

To improve the UAV's receive SINR, the conventional CB can be applied, i.e., all the available BSs in $\Omega$ cooperatively transmit the same message $x_u$ to the UAV simultaneously subject to their per-BS power constraints, so that their signals are in-phase at the UAV receiver and thus add constructively\cite{mudumbai2009distributed}. In this case, the UAV's receive SINR is re-expressed as
\begin{equation}\label{sinr2}
\gamma_{u,\text{CB}} = \frac{{\left(\sum\limits_{n \in \Omega}\lvert f_n \rvert\sqrt {P}\right)}^2}{\sigma^2 + P\sum\limits_{j \in {\cal J}_o} {\lvert f_j \rvert^2}}.
\end{equation}
By comparing (\ref{sinr1}) and (\ref{sinr2}), it is observed that applying CB is able to enhance the UAV's desired signal power and thus receive SINR, thanks to the beamforming gain by the cooperative transmission of all available BSs. However, such beamforming gain may be insufficient to overcome the strong aggregate terrestrial interference, if the number of available BSs is small, which is usually the case when the terrestrial UE density is high. In addition, increasing the BS transmit power $P$ can only marginally improve the UAV's SINR given in (\ref{sinr2}) since the signal power and the interference power both increase with $P$ in the same order. In view of the above limitations, the conventional CB scheme could be also ineffective to deal with the strong downlink interference to the UAV communication.

\subsection{Proposed CB with ITC}
A new CB scheme with ITC is proposed in this paper, in order to achieve the CB gains for both UAV signal enhancement (as in the conventional CB) and terrestrial interference suppression (via the newly proposed ITC). To characterize the optimal performance of the proposed scheme, we assume for the time being that the transmitted data symbols for all $K$ co-channel terrestrial UEs, i.e., $\{x_j\}_{j \in {\cal J}_o}$, are known at each of the available BSs serving the UAV, while we will consider the practical case with partial symbol knowledge and other implementation aspects of the proposed scheme in Section \ref{dis.CB}. Note that the proposed scheme can be extended to the case with only partial knowledge of $\{x_j\}_{j \in {\cal J}_o}$ at each serving BS of the UAV, by simply setting its transmit power for canceling the interference of a terrestrial UE with unknown symbol knowledge to zero. Under the proposed CB scheme with ITC, each available BS in general transmits the UAV's data symbol, along with all the co-channel terrestrial UEs' data symbols with different power allocations for ITC subject to its maximum power budget.

Let $w_{n,j}$ and $w_{n,u}$ be the complex beamforming weights used by BS $n \in \Omega$ to transmit the data symbol of the terrestrial UE served by BS $j, j \in {\cal J}_o$, and that of the UAV, respectively. As such, the received signal at the UAV becomes
\begin{equation}\label{comp1}
y_{u,{\text{CB-ITC}}}\!=\!\sum\limits_{n \in \Omega} {f_n\!\!\sum\limits_{j \in {\cal J}_o}\!\!{w_{n,j}x_j}}+\!\sum\limits_{n \in \Omega}\!\!{f_n{w_{n,u}}{x_u}}+\!\sum\limits_{j \in {\cal J}_o} \!\!{\sqrt{P}f_jx_j}+\!z_u.
\end{equation}
Obviously, to satisfy the total power constraint at each available BS $n \in \Omega$, we have
\begin{equation}\label{perBS1}
\sum\limits_{j \in {{\cal J}_o}} {{\lvert w_{n,j} \rvert}^2}  + {\lvert w_{n,u} \rvert^2} \le P, \forall n \in \Omega.
\end{equation}
By stacking the beamforming weights for different data symbols, we define ${\mv w}_j \triangleq [w_{n,j}]_{n \in \Omega} \in {\mathbb C}^{N}, j \in {\cal J}_o$ and ${\mv w}_u \triangleq [w_{n,u}]_{n \in \Omega} \in {\mathbb C}^{N}$. Then the received signal in (\ref{comp1}) can be more concisely expressed as
\begin{equation}\label{comp2}
y_{u,{\text{CB-ITC}}} = \sum\limits_{j \in {\cal J}_o} {({\mv f}_u^H{{\mv w}_j} + \sqrt{P}f_j)x_j} + {\mv f}_u^H{{\mv w}_u}{x_u} + {z_u},
\end{equation}
where ${\mv f}_u \triangleq [f_n^*]_{n \in \Omega} \in {\mathbb C}^{N}$. As a result, the UAV's receive SINR achievable by the proposed CB with ITC is given by
\begin{equation}\label{comp3}
\gamma_{u,{\text{CB-ITC}}} =\frac{{\lvert {{\mv f}_u^H{{\mv w}_u}} \rvert}^2}{\sigma ^2 + \sum\limits_{j \in {{\cal J}_o}} {{\lvert {{\mv f}_u^H{{\mv w}_j} + \sqrt{P}f_j} \rvert}^2}}.
\end{equation}
In addition, the per-BS power constraint in (\ref{perBS1}) is equivalent to
\begin{equation}\label{perBS2}
{\left[ {\sum\limits_{j \in {\cal J}_o} {{\mv w}_j}{\mv w}_j^H  + {{\mv w}_u}{\mv w}_u^H} \right]_i} \le P, \;i=1,2,\cdots,N,
\end{equation}
where $\left[ \cdot \right]_i$ denotes the $i$-th diagonal element of a square matrix.

It is worth noting that if there exist feasible solutions such that ${\mv f}_u^H{{\mv w}_j} + \sqrt{P}f_j = 0, \forall j \in {\cal J}_o$, then the UAV would be free from the terrestrial interference under the proposed CB scheme with ITC. As a result, increasing $P$ may yield more significant improvement of the UAV's receive SINR in the high transmit power regime as compared to the conventional CB without ITC. Besides, it is noted that the interference transmission by the serving BSs of the UAV would not cause additional interference to the existing co-channel terrestrial communications thanks to our assumed BS-UAV association rule given in (\ref{serv}).

\section{Problem Formulation}
With the proposed CB scheme with ITC, we aim to maximize the UAV's receive SINR as given in (\ref{comp3}) subject to the per-BS power constraints in (\ref{perBS2}). To this end, we need to jointly design the interference beamformers $\{{\mv w}_j\}_{j \in {\cal J}_o}$ and signal beamformer ${\mv w}_u$ at all available BSs (or serving BSs of the UAV) for transmitting the terrestrial UEs' messages and the UAV's message, respectively. The optimization problem is thus formulated as
\begin{align}
\text{(P1)} \mathop {\max}\limits_{\{{\mv w}_j\},{\mv w}_u}&\; \frac{{\lvert {\mv f}_u^H{{\mv w}_u} \rvert}^2}{\sigma ^2 + \sum\limits_{j \in {{\cal J}_o}} {{\lvert {\mv f}_u^H{{\mv w}_j} + \sqrt{P}f_j \rvert}^2}}\label{op1}\\
\text{s.t.}\;\;&{\left[ {\sum\limits_{j \in {\cal J}_o} {{\mv w}_j}{\mv w}_j^H  + {{\mv w}_u}{\mv w}_u^H} \right]_i} \le P, \;i=1,2,\cdots,N. \label{op1a}
\end{align}
Notice that by imposing ${\mv w}_j={\mv 0}, \forall j \in {\cal J}_o$ in (P1), the proposed CB scheme reduces to the conventional CB without ITC. Consequently, the solution to (P1) should generally yield a higher UAV SINR than the conventional CB without ITC.

Let $\{{\mv w}^\star_j\}$ and ${\mv w}^\star_u$ be an optimal solution to (P1). Then we can obtain the following proposition.
\begin{proposition}\label{opt.phase}
Given the amplitude of $\{{\mv w}^\star_j\}$ and ${\mv w}^\star_u$, their phases should satisfy
\begin{align}
\angle w_{n,u}^\star &= -\angle f_n, n \in \Omega,\label{opt.phase1}\\
\angle w_{n,j}^\star &= \angle f_j-\angle f_n + \pi, n \in \Omega, j \in {\cal J}_o.\label{opt.phase2}
\end{align}
\end{proposition}
\begin{IEEEproof}
Since the phases of $w_{n,u}^\star$ and $w_{n,j}^\star$'s will not affect the feasibility of power constraints in (\ref{op1a}), they can be freely chosen from $0$ to $2\pi$ to maximize (\ref{op1}). Notice that maximizing (\ref{op1}) can be decoupled into maximizing its numerator and minimizing its denominator. Based on this fact, we first show that its numerator can be maximized via (\ref{opt.phase1}). Specifically, the following must hold, i.e.,
\begin{equation}\label{ineq1}
\lvert {\mv f}_u^H{{\mv w}_u^\star} \rvert = \left| \sum\limits_{n \in \Omega} {f_nw_{n,u}^\star} \right| \le \sum\limits_{n \in \Omega} {\lvert f_n \rvert}{\lvert w_{n,u}^\star \rvert},
\end{equation}
where we have used the inequality $\left|\sum\nolimits_i{a_i}\right| \le \sum\nolimits_i {\lvert a_i \rvert}$ for complex numbers $a_i$'s.
The inequality in (\ref{ineq1}) holds at equality when all $f_n w_{n,u}^\star$'s have the same phase. Obviously, with (\ref{opt.phase1}), each $f_n w_{n,u}^\star$ has a zero phase and thus the equality holds. Next, we show that the denominator of (\ref{op1}) can be minimized via (\ref{opt.phase2}). Specifically, we have
\begin{equation}
0 \le \lvert {\mv f}_u^H{{\mv w}_j^\star} + \sqrt{P}f_j \rvert = \left|\sum\limits_{n \in \Omega} {f_nw_{n,j}^\star}+\sqrt{P}f_j \right|, j \in {\cal J}_o.
\end{equation}

It is easy to verify that
\begin{equation}\label{ineq2}
\sqrt{P}\lvert f_j \rvert \ge \sum\limits_{n \in \Omega} {\lvert f_n \rvert}{\lvert w_{n,j}^\star \rvert}
\end{equation}
must hold. Since otherwise, $\lvert w_{n,j}^\star\rvert$'s can be decreased until $\sqrt{P}\lvert f_j \rvert = \sum\nolimits_{n \in \Omega} {\lvert f_n \rvert}{\lvert w_{n,j}^\star \rvert}$ holds, under which we have $\lvert {\mv f}_u^H{{\mv w}_j^\star} + \sqrt{P}f_j \rvert=0$ by following (\ref{opt.phase2}), thus resulting in an objective value of (P1) that is no smaller than the previous one. Given (\ref{ineq2}), the following inequalities hold,
\begin{align}
\left|\sum\limits_{n \in \Omega} {f_nw_{n,j}^\star}+\sqrt{P}f_j \right| &\ge \sqrt{P}\lvert f_j \rvert - \left|\sum\limits_{n \in \Omega} {f_nw_{n,j}^\star}\right|\nonumber\\
&\ge \sqrt{P}\lvert f_j \rvert \!-\! \sum\limits_{n \in \Omega} {\lvert f_n \rvert}{\lvert w_{n,j}^\star \rvert} \ge 0, j \in {\cal J}_o,\label{ineq3}
\end{align}
where the first inequality is due to $\left|b+\sum\nolimits_i{a_i}\right| \ge \left|b-\lvert \sum\nolimits_i {a_i} \rvert\right|$ for complex numbers $b$ and $a_i$'s, while the second inequality is similar to that in (\ref{ineq1}). In order for the equalities in these two inequalities to hold at the same time, it is required that all $f_n w_{n,j}^\star$'s are of the negative sign of $f_j$, thus leading to (\ref{opt.phase2}). The proof is thus completed.
\end{IEEEproof}

Proposition \ref{opt.phase} reveals that we can find an optimal solution to (P1) if the optimal amplitude of $w_{n,u}^\star$'s and $w_{n,j}^\star$'s is obtained. In particular, (\ref{opt.phase1}) ensures that the UAV's received signals from all available BSs are in-phase and constructively combined. On the other hand, (\ref{opt.phase2}) shows that the ITC signals from all available BSs are in opposite phase to their corresponding terrestrial interference at the UAV receiver so as to maximally cancel it. In addition, due to (\ref{ineq2}), the terrestrial interference is ensured to be reduced.

In order to obtain the amplitude of $w_{n,u}^\star$'s and $w_{n,j}^\star$'s, we substitute (\ref{opt.phase1}) and (\ref{opt.phase2}) into (P1), and the complex beamformer design in (P1) is reduced to power allocations at each available BS for transmitting $x_u$ and $x_j$'s. For convenience, we define ${\mv v}_j \triangleq [\lvert w_{n,j} \rvert]_{n \in \Omega} \in {\mathbb R}^{N}, j \in {\cal J}_o$, ${\mv v}_u \triangleq [\lvert w_{n,u} \rvert]_{n \in \Omega} \in {\mathbb R}^{N}$ and ${\mv h}_u \triangleq [\lvert f_n \rvert]_{n \in \Omega} \in {\mathbb R}^{N}$, by stacking the amplitude of the beamforming weights for different data symbols over all available BSs and those of the channel gains from all available BSs to the UAV, respectively. Then, (P1) is simplified to the following problem,
\begin{align}
\text{(P2)} \mathop {\max}\limits_{\{{\mv v}_j\},{\mv v}_u}&\; \frac{({\mv h}_u^T{\mv v}_u)^2}{\sigma ^2 + \sum\limits_{j \in {{\cal J}_o}} {(\sqrt{P}\lvert f_j \rvert-{\mv h}_u^T{\mv v}_j)^2}}\label{op2}\\
\text{s.t.}\;\;&{\left[ {\sum\limits_{j \in {\cal J}_o} {{\mv v}_j}{\mv v}_j^T  + {{\mv v}_u}{\mv v}_u^T} \right]_i} \le P, \;i=1,2,\cdots,N, \label{op2a}\\
&{\mv v}_u \succeq {\bf 0}, {\mv v}_j \succeq {\bf 0}, \forall j \in {\cal J}_o.\label{op2b}
\end{align}

However, (P2) can be shown to be a non-convex optimization problem, and thus it is difficult to solve (P2) optimally in its current form. Moreover, there is an intricate and non-trivial trade-off in the power allocations for maximizing the UAV's received signal power (see the numerator of (\ref{op2})) versus minimizing the residual terrestrial interference power after ITC (in the denominator of (\ref{op2})), subject to their total transmit power constraints in (\ref{op2a}). For example, if more power is assigned to increase the former, less terrestrial interference will be cancelled in general and thus the latter may be increased too. As a result, both the numerator and the denominator in (\ref{op2}) may increase and it is unclear whether the UAV's receive SINR will increase or not. Similarly, if more power is assigned to cancel the terrestrial interference, the latter decreases in general and so does the former, and as such it is also unknown whether increment or decrement in the UAV's receive SINR will be resulted. Therefore, both ${\mv v}_j$'s and ${\mv v}_u$ have non-trivial effects on the UAV's receive SINR or the objective value of (P2). To draw useful insights into them, in the next section, we first derive the optimal solution to problem (P2) in the special case with one serving (available) BS of the UAV, i.e., $N=1$, then solve (P2) for the general case with $N>1$.

\section{Optimal Design of CB with ITC}
In this section, we solve (P2) for the cases of $N=1$ and $N>1$, respectively.

\subsection{Special Case with $N=1$}\label{single.av}
First, we consider the special case of (P2) with one single ($N=1$) available BS serving the UAV and derive the closed-form optimal solution to it. To start with, we consider a further simplified scenario with one single occupied BS with $K=1$, i.e., only one terrestrial interferer (or co-channel UE) with the UAV. We then extend the solution to the case with $K \ge 2$. Denote by $f_a$ ($f_o$) the channel coefficient from the single available (occupied) BS to the UAV. With $N=1$ and $K=1$, we only need to determine the power allocations at the available BS for transmitting the UAV's and the single terrestrial UE's data symbols, denoted as $v_j$ and $v_u$, respectively. As such, problem (P2) is simplified as
\begin{align}
\nonumber \text{(P3)}\; \mathop {\max}\limits_{v_j,v_u \ge 0}&\; \frac{\lvert f_a \rvert^2v^2_u}{\sigma ^2 + (\lvert f_o \rvert\sqrt{P} -\lvert f_a \rvert v_j)^2}\nonumber\\
\text{s.t.}\;\;&v_j^2  + v_u^2 \le P.\label{op3}
\end{align}

It is easy to verify that at the optimality of (P3), we must have $\lvert f_o \rvert\sqrt{P} -\lvert f_a \rvert v_j \ge 0$, i.e., the terrestrial interference is ensured to be reduced, since otherwise we can simultaneously decrease $v_j$ and increase $v_u$ to attain a larger objective value. By following the similar procedures as will be given later in Section \ref{opt.socp}, problem (P3) can be shown to be essentially a convex optimization problem that also satisfies the Slater's condition\cite{boyd2009convex}. As such, the strong duality holds between problem (P3) and its dual problem, and their optimal solutions should satisfy the Karush-Kuhn-Tucker (KKT) conditions. Based on the KKT conditions of (P3), we can obtain the optimal solution of (P3) in closed form, as given in the following proposition.
\begin{proposition}\label{single.opt}
The optimal solution to (P3), denoted by $(v_j^\star,v_u^\star)$, is given by
\begin{align}
v_j^\star&=\frac{X - \sqrt {X^2 - 4{\lvert f_a \rvert}^2{\lvert f_o \rvert}^2P^2}}{2{\lvert f_a \rvert}{\lvert f_o \rvert}\sqrt P},\label{solution1}\\
v_u^\star&=\sqrt {P - v_j^{\star 2}},\label{solution2}
\end{align}
where $X \triangleq \sigma^2 + ({\lvert f_a \rvert}^2 + {\lvert f_o \rvert}^2)P$. Moreover, the UAV's maximum receive SINR, denoted as $\eta^\star$, is given by
\begin{equation}\label{opt.eta}
\eta^\star=\frac{-Y + \sqrt {Y^2 + 4\sigma^2P{{\lvert f_a \rvert}^2}}}{2\sigma ^2},
\end{equation}
with $Y \triangleq \sigma ^2 + ({\lvert f_o \rvert}^2 - {\lvert f_a \rvert}^2)P$.
\end{proposition}
\begin{IEEEproof}
Please refer to the appendix.
\end{IEEEproof}

In contrast, if the available BS only transmits the UAV's symbol without ITC, i.e., $v_j=0$ and $v_u=\sqrt{P}$, then the resultant SINR, denoted as $\eta_0$, is given by
\begin{equation}\label{SINRnoif}
{\eta_0}=\frac{P{\lvert f_a \rvert}^2}{\sigma^2+P{\lvert f_o \rvert}^2}.
\end{equation}

Let $\rho_j=\frac{v_j^{\star 2}}{P}$ and $\rho_u=\frac{v_u^{\star 2}}{P}=1-\rho_j$ be the optimal power allocation ratios for ITC and the UAV's desired signal, respectively. According to Proposition \ref{single.opt}, we obtain the following proposition.
\begin{proposition}\label{pw.ratio}
$\rho_j$ is a monotonically increasing function of the BS transmit power $P$. Thus, $\rho_u$ is a monotonically decreasing function of $P$.
\end{proposition}
\begin{IEEEproof}
Since $\rho_u=1-\rho_j$, we only need to prove that $\rho_j$ is monotonically increasing with $P$. Notice that $\rho_j = \left( \frac{v_j^\star}{\sqrt P} \right)^2$ and $v_j^\star>0$. Consequently, it suffices to prove that $\tilde\rho_j \triangleq \frac{v_j^\star}{\sqrt P}$ is monotonically increasing with $P$. It is easy to verify that the first derivative of $\tilde\rho_j$ satisfies
\begin{equation}
\frac{{\text{d}}{\tilde\rho}_j}{{\text{d}}P} = \frac{v_j^\star\sigma^2}{P\sqrt {{\left(\sigma ^2+{\lvert f_a \rvert}^2P + {\lvert f_o \rvert}^2P\right)}^2 - 4{\lvert f_a \rvert}^2{\lvert f_o \rvert}^2P^2}} > 0.
\end{equation}
Therefore, $\rho_j$ is monotonically increasing with $P$. The proof is thus completed.
\end{IEEEproof}

Proposition \ref{pw.ratio} reveals that as $P$ increases, more transmit power should be allocated for ITC. As a result, the proposed CB scheme with ITC can yield higher SINR than that given in (\ref{SINRnoif}) by the conventional CB without ITC. In particular, we can derive the following asymptotic results under the two extreme cases of $P \to 0$ and $P \to \infty$.
\begin{lemma}\label{limit1}
When $P \to 0$, we have ${\eta^\star}={\eta_0}=0$, $\rho_j=0$, and $\rho_u=1$.
\end{lemma}
\begin{lemma}\label{limit2}
When $P \to \infty$, the following limits hold:
\begin{align}
&\rho_j=
\begin{cases}
\frac{\lvert f_o \rvert^2}{\lvert f_a \rvert^2}, &\text{if}\;{\lvert f_a \rvert}^2 \ge {\lvert {f_o} \rvert}^2\\
\frac{\lvert f_a \rvert^2}{\lvert f_o \rvert^2}, &\text{otherwise,}
\end{cases}, \label{rho.j}\\
&{\eta^\star}=
\begin{cases}
\frac{P({\lvert f_a \rvert}^2 - {\lvert f_o \rvert}^2)}{\sigma ^2}\to \infty &\text{if}\;{\lvert f_a \rvert}^2 > {\lvert f_o \rvert}^2\\
\frac{{\lvert f_a \rvert \sqrt{P}}}{\sigma}\to \infty &\text{if}\;{\lvert f_a \rvert}^2 = {\lvert {f_o} \rvert}^2\\
\frac{{\lvert f_a \rvert}^2}{{\lvert f_o \rvert}^2-{\lvert f_a \rvert}^2}, &\text{otherwise,}
\end{cases},\\
&{\eta_0}=\frac{{\lvert f_a \rvert}^2}{{\lvert f_o \rvert}^2}.
\end{align}
\end{lemma}

Lemma \ref{limit1} reveals that in the low transmit power regime (i.e., $P \to 0$), the transmit power should be all assigned to the UAV's message. This is expected as the interference in this case is much weaker than the receiver noise, which implies that the performance gain of the proposed CB scheme over the conventional CB without ITC is small and diminishes as $P \to 0$. On the other hand, Lemma \ref{limit2} reveals that the proposed scheme significantly outperforms the conventional CB in the high transmit power regime (i.e, $P \rightarrow \infty$) or equivalently the high signal-to-noise ratio (SNR) regime, which is typically the case for practical UAV downlink communications in cellular networks due to the LoS-dominant channels with BSs. Moreover, the power allocation ratio for ITC finally converges to a fixed value given in (\ref{rho.j}). In particular, if ${\lvert f_a \rvert}^2 \ge {\lvert {f_o} \rvert}^2$, we have $\rho_j=\frac{\lvert f_o \rvert^2}{\lvert f_a \rvert^2}$, resulting in $\lvert f_o \rvert\sqrt{P} -\lvert f_a \rvert v_j^\star = \lvert f_o \rvert\sqrt{P} -\lvert f_a \rvert\sqrt{P\rho_j} = 0$, i.e., the terrestrial interference should be completely cancelled in this case; otherwise, if ${\lvert f_o \rvert}^2 > {\lvert {f_a} \rvert}^2$, then the terrestrial interference should be partially canceled.

\textbf{Numerical Example}: To further characterize the behavior of the optimal value of (P3) or the maximum UAV receive SINR under different BS transmit power, we provide the following two numerical examples. Define $\gamma_1 \triangleq \frac{\lvert f_a \rvert^2v_u^{\star 2}}{\sigma^2}$ as the ratio of the UAV's received signal power to the noise power, and $\gamma_2 \triangleq \frac{\sigma ^2 + ({\lvert f_a \rvert v_j^\star-\lvert f_o \rvert\sqrt{P}})^2}{\sigma^2}$ as the ratio of the UAV's noise-plus-residual-interference (after ITC) power to the noise power; thus, $\eta^\star=\frac{\gamma_1}{\gamma_2}$. In the first example, we set ${\lvert f_a \rvert^2}/{\sigma^2}=10$ and ${\lvert f_o \rvert^2}/{\sigma^2}=12$. While in the second example, we set ${\lvert f_a \rvert^2}/{\sigma^2}=15$ and ${\lvert f_o \rvert^2}/{\sigma^2}=12$.

\begin{figure}[hbtp]
\centering
\subfigure[$\gamma_1$, $\gamma_2$, $\eta^\star$ and $\eta_0$ versus $P$.]{\includegraphics[width=0.4\textwidth]{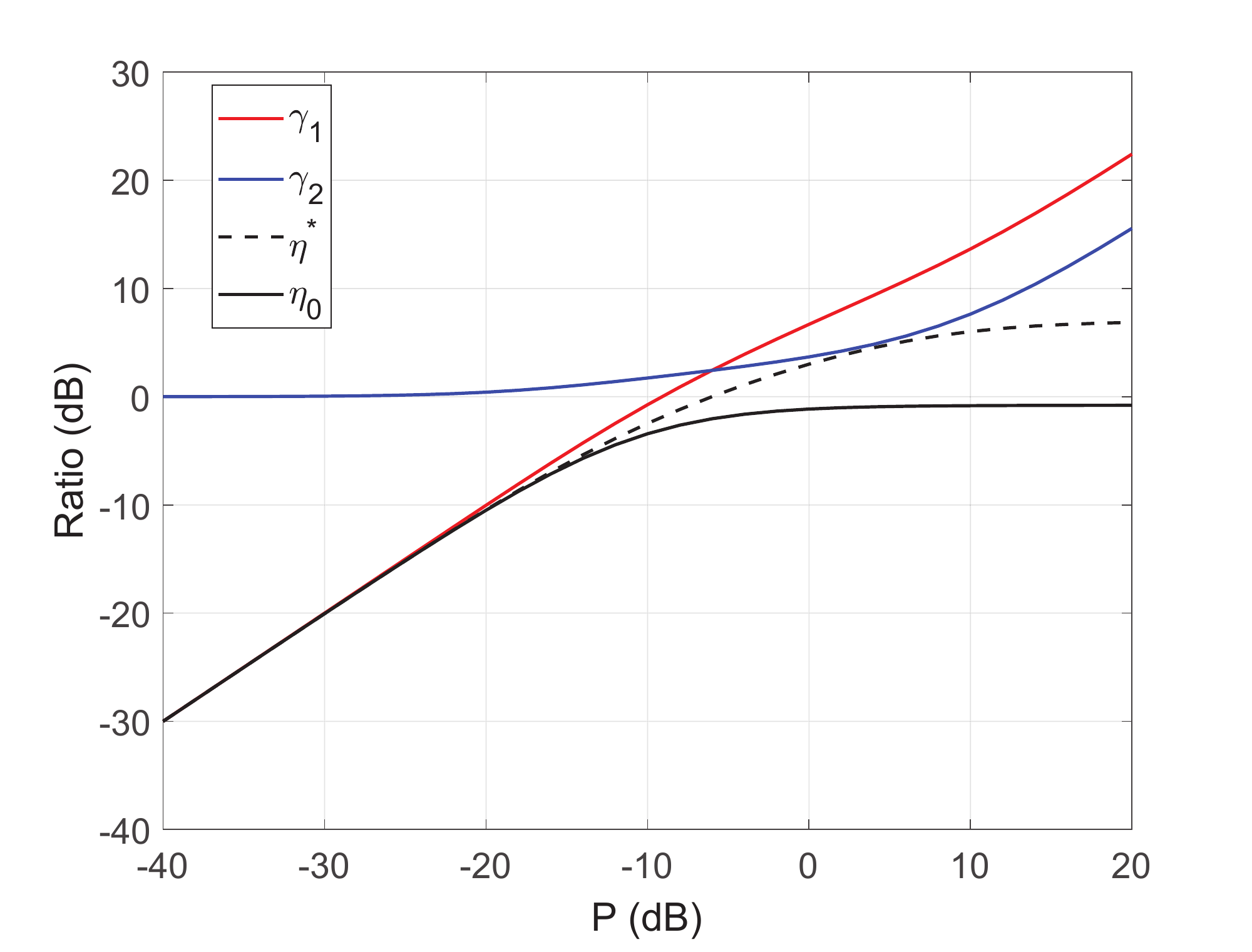}}
\subfigure[Power allocation ratio versus $P$.]{\includegraphics[width=0.4\textwidth]{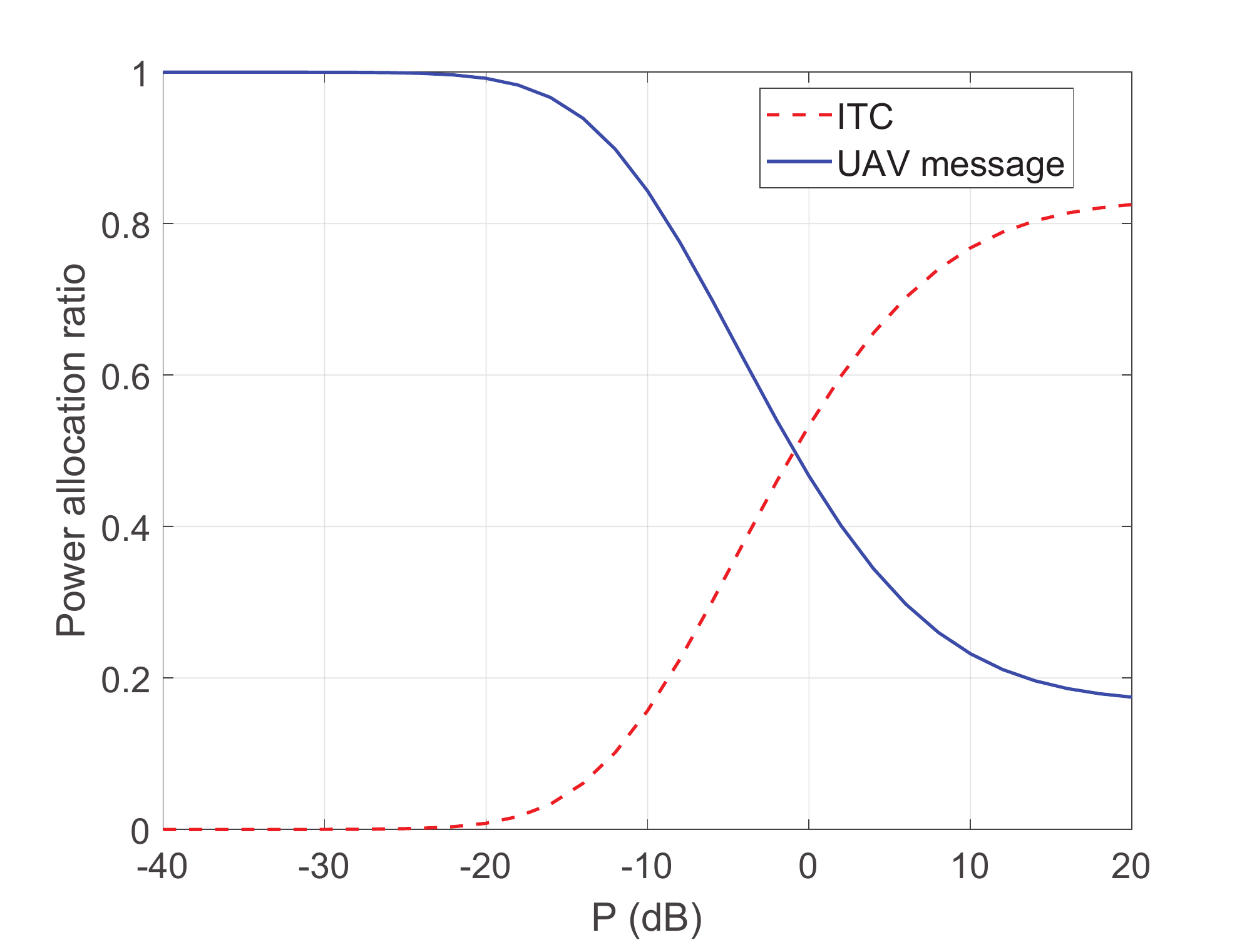}}
\caption{Results of example 1 for (P3).}\label{Grp1}
\end{figure}
First, Fig.\,\ref{Grp1}(a) shows $\gamma_1$, $\gamma_2$, $\eta^\star$, and $\eta_0$ (all in dB) versus the BS transmit power $P$ for example 1. It is observed that in the low-to-medium transmit power regime, $\gamma_1$ increases much faster than $\gamma_2$ with $P$. As a result, $\eta^\star$ also increases rapidly with $P$. This is because the terrestrial interference is not dominant over the noise in this regime. However, in the high transmit power regime, since ${\lvert f_a \rvert}^2 < {\lvert f_o \rvert}^2$, the terrestrial interference should not be completely cancelled in this example, and the residual interference dominates over the noise, leading to comparable increasing rates between $\gamma_1$ and $\gamma_2$. Thus, $\eta^\star$ increases marginally with $P$ in the high transmit power regime, and finally converges to a finite value given in Lemma \ref{limit2}. In addition, in Fig.\,\ref{Grp1}(b), we plot $\rho_j$ and $\rho_u$ versus $P$. It is observed that in the medium transmit power regime, with the increasing residual terrestrial interference with $P$, the power allocation ratio for ITC increases notably. By contrast, in the high transmit power regime, the power allocation ratio for ITC increases at a slower rate and finally converges to the limit value of ${\lvert f_a \rvert^2}/{\lvert f_o \rvert^2}=10/12=0.83$, in accordance with Lemma \ref{limit2}.

\begin{figure}[hbtp]
\centering
\subfigure[$\gamma_1$, $\gamma_2$, $\eta^\star$ and $\eta_0$ versus $P$.]{\includegraphics[width=0.4\textwidth]{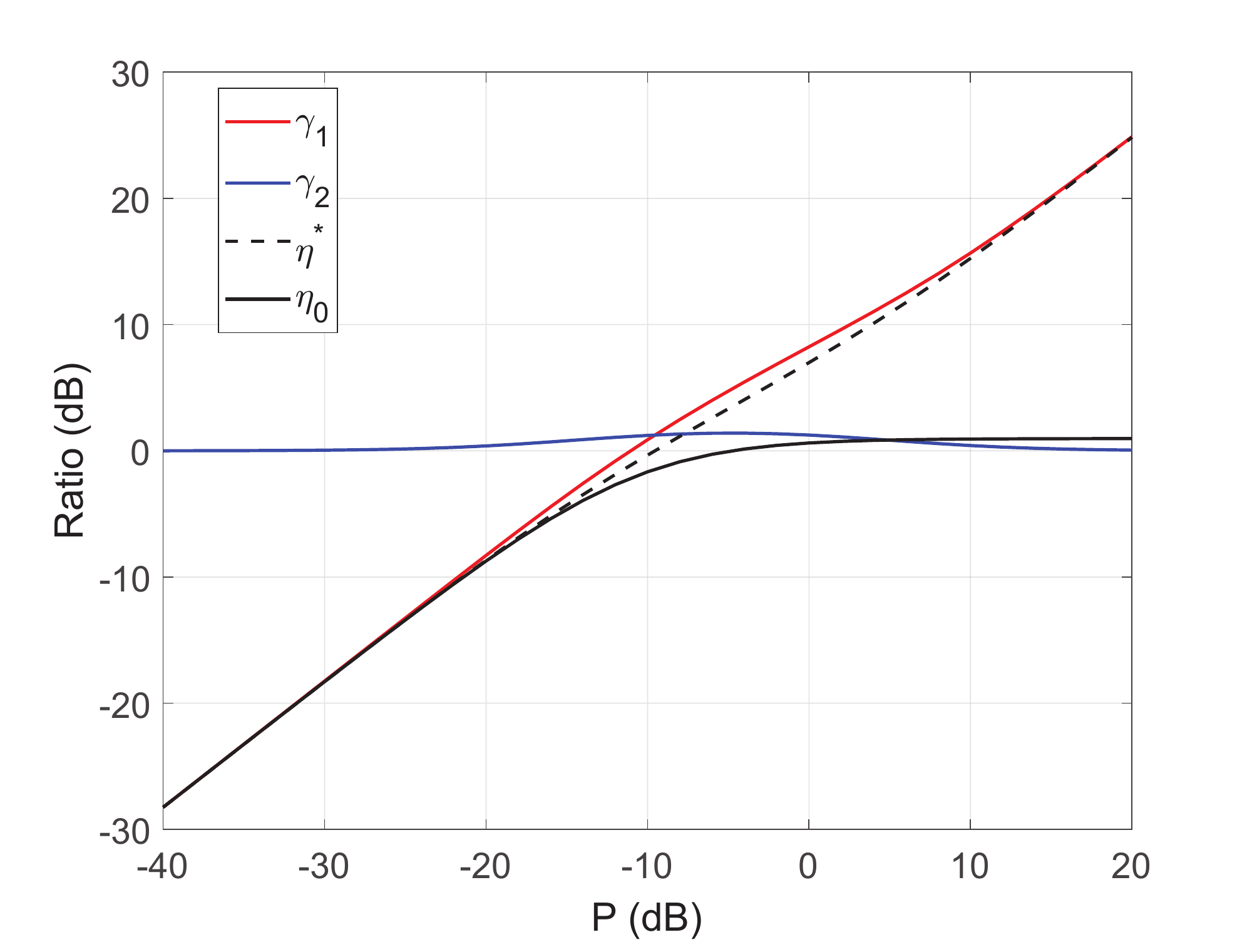}}
\subfigure[Power allocation ratio versus $P$.]{\includegraphics[width=0.4\textwidth]{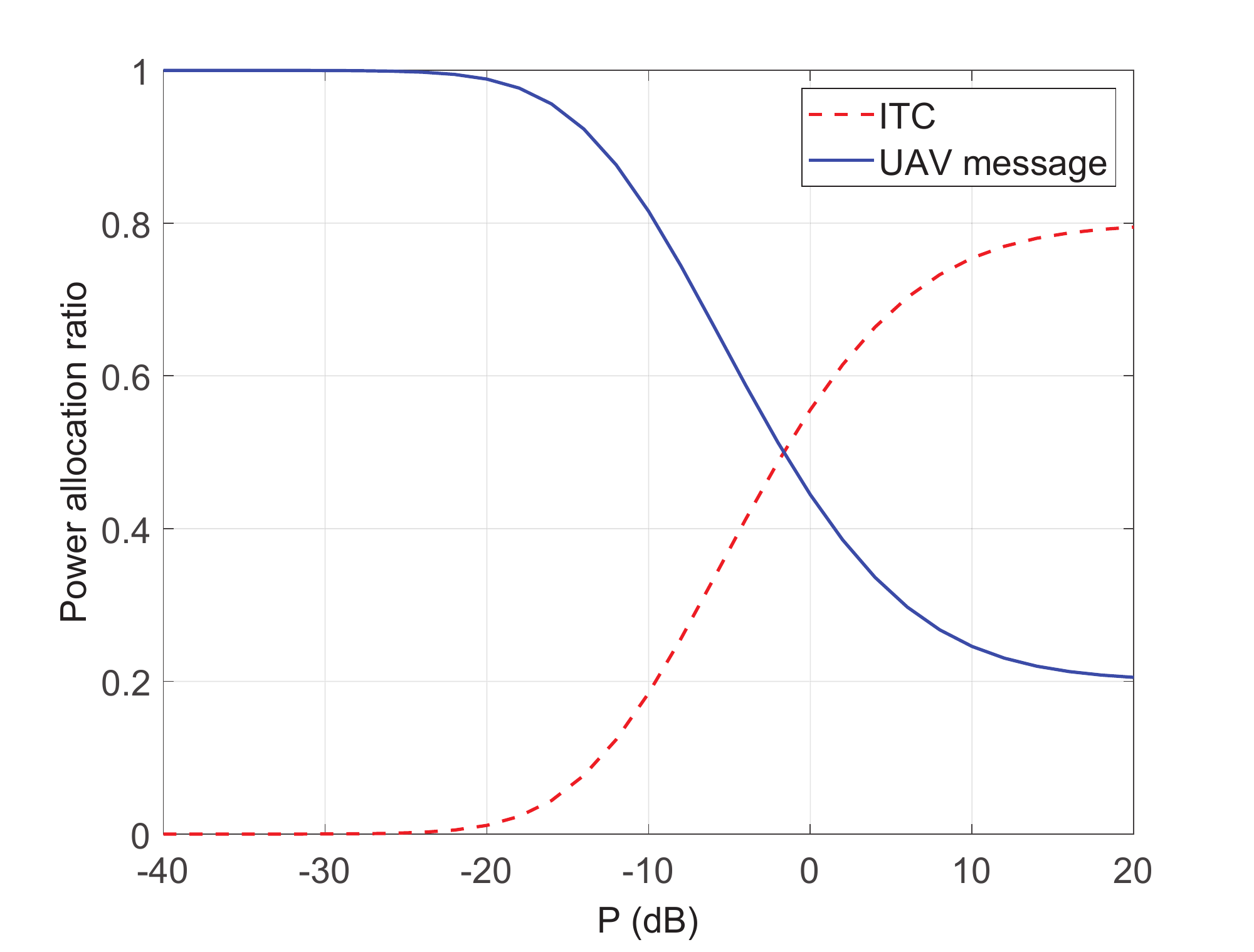}}
\caption{Results of example 2 for (P3).}\label{Grp2}
\end{figure}
Next, we plot the results of example 2 with $\lvert f_a \rvert^2 > \lvert f_o \rvert^2$ in Fig.\,\ref{Grp2}. In contrast to Fig.\,\ref{Grp1}(a), it is observed from Fig.\,\ref{Grp2}(a) that the residual terrestrial interference is consistently below a low level over the whole range of $P$. As a result, $\eta^\star$ is observed to increase rapidly with $P$ even in the high transmit power regime. The main reason lies in that with ${\lvert f_a \rvert}^2 > {\lvert f_o \rvert}^2$ in this example, the available BS has the ability to completely cancel the terrestrial interference, which makes the residue interference comparable to the receiver noise. From Fig.\,\ref{Grp2}(b), it is observed that the power allocation ratios in example 2 behave similarly to those in example 1.

Next, we present the optimal solution to (P2) in the case with $N=1$ and $K \ge 2$, for which the derivations are similar to the case of $K=1$ shown above and thus omitted for brevity. In this case, we need to optimize the power allocations at the available BS among the UAV's and more than one terrestrial UE's data symbols, denoted as $v_u$ and $\{v_j\}_{j \in {\cal J}_o}$, respectively. The closed-form optimal solutions are presented in the following proposition.
\begin{proposition}\label{weight}
When $N=1$ and $K \ge 2$, the optimal solution to (P2) is given by
\begin{align}
v_j^\star&=\frac{\eta^\star}{\eta^\star + 1}\frac{\lvert f_j\rvert\sqrt P}{\lvert f_a \rvert}, j \in {\cal J}_o,\label{solution3}\\
v_u^\star&=\sqrt {P - \sum\limits_{j \in {\cal J}_o} {v_j^{\star2}}},\label{solution4}
\end{align}
\end{proposition}
where $\eta^\star$ is the UAV's maximum receive SINR given by
\begin{equation}
\eta^\star=\frac{-\sigma ^2 + PS_a + \sqrt{(\sigma ^2 - PS_a)^2 + 4\sigma^2P{\lvert f_a \rvert^2}}}{2\sigma ^2},
\end{equation}
with $S_a \triangleq {\lvert f_a \rvert}^2 - \sum\nolimits_{j \in {\cal J}_o}{\lvert {f_j} \rvert}^2$ denoting the difference between the channel power gain from the available BS and the sum channel power gain from all occupied BSs.

From (\ref{solution3}), it is noted that the optimal power allocations to the terrestrial UEs' symbols (i.e., $\rho_j$'s) are proportional to their corresponding occupied (serving) BSs' interference channel power gains with the UAV, i.e., $\lvert f_j \rvert^2, j \in {\cal J}_o$. Moreover, we have
\begin{equation}\label{ITC.cap}
\frac{{\text{d}}{\eta^\star}}{{\text{d}}S_a} = \frac{P\eta^\star}{\sqrt{(\sigma ^2 - PS_a)^2 + 4\sigma^2P{\lvert f_a \rvert^2}}}>0.
\end{equation}
The above result implies that the UAV's maximum receive SINR is an increasing function of $S_a$, which thus determines the ITC performance of the available BS serving the UAV. This observation will play an important role in the design of distributed CB with ITC in Section \ref{dis.CB}.\vspace{-8pt}

\subsection{Optimal Solution with $N>1$}\label{opt.socp}
For the general case with $N>1$, it is difficult to obtain the optimal solution to (P2) in closed form. Nonetheless, the problem can be numerically solved by recasting (P2) as a convex second-order cone programming (SOCP) optimization problem. To this end, we first introduce a slack variable $b$ and rewrite (P2) as
\begin{subequations}\label{op4}
\begin{align}
\nonumber \mathop {\max}\limits_{\{{\mv v}_j\},{\mv v}_u,b}&\; \frac{({{\mv h}_u^T{{\mv v}_u}})^2}{b^2}\nonumber\\
\text{s.t.}\;\;&\sigma ^2 + \sum\limits_{j \in {{\cal J}_o}} {(\sqrt{P}\lvert f_j \rvert-{\mv h}_u^T{\mv v}_j)^2} \le b^2, \label{op4a}\\
&{\left[ {\sum\limits_{j \in {\cal J}_o} {{\mv v}_j}{\mv v}_j^T + {{\mv v}_u}{\mv v}_u^T} \right]_i} \le P, \;i=1,2,\cdots,N, \label{op4b}\\
&b \ge 0, {\mv v}_u \succeq {\bf 0}, {\mv v}_j \succeq {\bf 0}, \forall j \in {\cal J}_o.\label{op4c}
\end{align}
\end{subequations}
It is easy to verify that at the optimality of problem (\ref{op4}), constraint (\ref{op4a}) must hold with equality, since otherwise we can decrease $b$ to attain a larger objective value. This validates that problem (\ref{op4}) is equivalent to (P2).

Next, we introduce the following variable transformation:
\begin{equation}
\tilde {\mv v}_u = \frac{{\mv v}_u}{b}, \;\tilde {\mv v}_j = \frac{{\mv v}_j}{b}, \forall j \in {{\cal J}_o}, \; \tilde b = \frac{1}{b}.
\end{equation}
Consequently, problem (\ref{op4}) can be reformulated as
\begin{subequations}\label{op5}
\begin{align}
\nonumber \mathop {\max}\limits_{\{\tilde{\mv v}_j\},\tilde {\mv v}_u,\tilde b}&\; ({\mv h}_u^T{\tilde{\mv v}_u})^2\nonumber\\
\text{s.t.}\;\;&\sigma ^2 {\tilde b}^2 + \sum\limits_{j \in {{\cal J}_o}} {(\sqrt{P}\lvert f_j \rvert{\tilde b}-{{\mv h}_u^T{\tilde{\mv v}_j}})^2} \le 1, \label{op5a}\\
&{\left[ {\sum\limits_{j \in {\cal J}_o} {\tilde{\mv v}_j}\tilde{\mv v}_j^T + {\tilde{\mv v}_u}\tilde{\mv v}_u^T} \right]_i} \le P{\tilde b}^2, \;i=1,2,\cdots,N, \label{op5b}\\
&\tilde b \ge 0, \tilde{\mv v}_u \succeq {\bf 0}, \tilde{\mv v}_j \succeq {\bf 0}, \forall j \in {\cal J}_o.\label{op5c}
\end{align}
\end{subequations}

Notice that both constraints (\ref{op5a}) and (\ref{op5b}) can be rewritten as SOCP constraints given by
\begin{equation}\label{socp}
\begin{split}
&\left\| {\begin{array}{*{20}{c}}
{\left[\sqrt{P}\lvert f_j \rvert{\tilde b}-{{\mv h}_u^T{\tilde{\mv v}_j}}\right]_{j \in {\cal J}_o}}\\
{\sigma \tilde b}
\end{array}} \right\|  \le 1,\\
&\left\| {\begin{array}{*{20}{c}}
{\left[{\mv e}_i^T\tilde{\mv v}_j\right]_{j \in {\cal J}_o}}\\
{{\mv e}_i^T\tilde{\mv v}_u}
\end{array}} \right\| \le \sqrt P \tilde b, \;i=1,2,\cdots,N,
\end{split}
\end{equation}
respectively, where ${\mv e}_i$ denotes the $i$-th column of an identity matrix ${\mv I}_N$. However, problem (\ref{op4}) is still non-convex, since the objective function of problem (\ref{op5}) is convex (instead of concave) in ${\tilde{\mv v}_u}$. Fortunately, since ${\mv h}_u \succeq \mathbf{0}$ and $\tilde{\mv v}_u \succeq \mathbf 0$, we have ${\mv h}_u^T{\tilde{\mv v}_u} \ge 0$. Thus, it is equivalent to maximizing ${\mv h}_u^T{\tilde{\mv v}_u}$ in problem (\ref{op5}). As a result, the objective function of problem (\ref{op5}) becomes affine in $\tilde{\mv v}_u$; hence, problem (\ref{op5}) is a convex optimization problem, which can be optimally solved via standard convex optimization techniques, e.g., the interior-point method or the off-the-shelf solver, \texttt{CVX}\cite{boyd2009convex}.

\section{Divide-and-Conquer Based Distributed CB Design}\label{dis.CB}
The optimal CB design with ITC solution presented in the previous section requires centralized implementation. Specifically, a central scheduler (e.g., one selected from the available BSs) needs to collect the channel state information (CSI) with the UAV from all the available and occupied BSs involved and compute the optimal CB solution, which is then sent to the available BSs for their downlink communication with the UAV. In addition, all occupied BSs need to send their terrestrial UEs' messages to the corresponding available BSs (with positive power allocations) for ITC. Such a centralized design achieves the optimal performance for the proposed scheme, but requires exorbitant CSI/message backhaul transmissions among the involved BSs, which may be practically costly to implement. To tackle this issue, in this section, we propose a distributed algorithm with significantly lower complexity and overhead than the centralized scheme for implementing the proposed CB with ITC more cost-effectively, based on a novel divide-and-conquer approach.

\subsection{Divide-and-Conquer Approach}
Different from the centralized CB, the proposed distributed CB only requires local backhaul transmissions between the occupied (co-channel) BSs and their neighboring available BSs (or serving BSs of the UAV). Specifically, each occupied BS $j \in {\cal J}_o$ only shares its served terrestrial UE's data symbol $x_j$ with the available BSs in its first $M$-tier neighborhood, denoted as $\Omega_j={\cal N}_j(M) \cap \Omega$, for ITC, instead of all available BSs in $\Omega$ as required in the centralized CB. For convenience, we refer to $M$ as the ``cooperation size'' in the sequel. Notice that under the terrestrial ICIC scheme considered in Section \ref{criterion}, there is no available BS in the first $q$-tier neighborhood of any occupied BS. Hence, the cooperation size $M$ should satisfy $M \ge q+1$ for implementing the proposed distributed CB. If $\Omega_j \ne \emptyset$, the residual terrestrial interference from occupied BS $j \in {\cal J}_o$ at the UAV receiver due to the ITC of the available BSs in $\Omega_j$ can be expressed as
\begin{equation}\label{resIF1}
I_j = \sum\limits_{n \in \Omega_j} {w_{n,j}}f_n + \sqrt P{f_j}.
\end{equation}
Based on (\ref{resIF1}), we propose a novel divide-and-conquer approach for distributed CB design with ITC, detailed as follows.

\textbf{``Divide'' part}: Each occupied BS $j$ splits its interference to the UAV (or equivalently its channel coefficient with the UAV, $f_j$) into $\lvert \Omega_j \rvert$ portions, i.e., $f_j = \sum\nolimits_{n \in \Omega_j} {f_{j,n}}$, with $f_{j,n} = f_j\theta_{j,n}$, $0 \le \theta_{j,n} \le 1, n \in \Omega_j$ and $\sum\nolimits_{n \in \Omega_j} \theta_{j,n}=1$. Accordingly, the residual interference in (\ref{resIF1}) can be re-expressed as
\begin{equation}\label{resIF2}
I_j = \sum\limits_{n \in \Omega_j} {(w_{n,j}{f_n} + \sqrt P {f_j\theta_{j,n}})}, j \in {\cal J}_o.
\end{equation}
It follows from (\ref{resIF2}) that each terrestrial interference portion of occupied BS $j$ (i.e., $f_{j,n}$) can be cancelled by a different available BS $n$ in $\Omega_j$. As such, all available BSs are able to perform the ITC in parallel, thus reducing the signaling overhead and overall delay.

\textbf{``Conquer'' part}: According to the definition of $\Omega_j$, each available BS $n \in \Omega$ only exchanges information with the occupied BSs in the set ${\cal J}_{o,n} \triangleq \left\{j\left| n \in \Omega_j \right.\right\}$, and cancels the terrestrial interference portions assigned to it, i.e., $f_{j,n}=f_j\theta_{j,n}, j \in {\cal J}_{o,n}$. To cancel them with its best effort, the available BS $n$ determines its power allocations $\{v_{n,j}\}_{j \in {\cal J}_{o,n}}$ and $v_{n,u}$ (i.e., the amplitude of its beamforming weights $\{w_{n,j}\}_{j \in {\cal J}_{o,n}}$ and $w_{n,u}$, respectively) according to Proposition \ref{weight}, by replacing $f_j$, $f_a$ and ${\cal J}_o$ with $f_j\theta_{j,n}$, $f_n$ and ${\cal J}_{o,n}$, respectively. The phases of its beamforming weights $\{w_{n,j}\}_{j \in {\cal J}_{o,n}}$ and $w_{n,u}$ are determined according to Proposition \ref{opt.phase}, by replacing ${\cal J}_o$ with ${\cal J}_{o,n}$. However, such a decentralized ITC approach fails to mitigate the UAV receiver noise optimally as in the centralized design due to distributed power allocations at the available BSs. Consequently, each available BS determines its power allocations with the full noise power $\sigma^2$ as in Proposition \ref{weight} by assuming that it is the sole serving BS of the UAV. This evidently exaggerates the effect of receiver noise and results in suboptimal ITC performance. To eliminate the noise effect and simplify the power allocations at available BSs, we consider the high SNR case with $P \gg \sigma^2$ in Proposition \ref{weight}. Then, the resultant power allocations at each available BS $n \in \Omega$ are given by
\begin{align}
v_{n,j}&=
\begin{cases}
\frac{\lvert f_j\rvert\sqrt P\theta_{j,n}}{\lvert f_n \rvert}, &{\text{if}}\;S_n \ge 0\\
\frac{\lvert f_n \rvert\lvert f_j\rvert\sqrt P\theta_{j,n}}{\sum\nolimits_{j \in {\cal J}_{o,n}}{\lvert f_j \rvert^2\theta_{j,n}^2}}, &\text{otherwise}
\end{cases}, j \in {\cal J}_{o,n},\label{solution5}\\
v_{n,u}&=\sqrt {P - \sum\limits_{j \in {\cal J}_{o,n}}{v_{n,j}^2}},\label{solution6}
\end{align}
where $S_n \triangleq {\lvert f_n \rvert}^2-\sum\nolimits_{j \in {\cal J}_{o,n}} {{\lvert f_j \rvert}^2\theta_{j,n}^2}$ indicates the ITC capability of available BS $n$, as shown in (\ref{ITC.cap}). From (\ref{solution5}), we have
\begin{equation}\label{res.int}
\sqrt P\theta_{j,n}\lvert f_j \rvert - \lvert f_n \rvert v_{n,j} =
\begin{cases}
0, &{\text{if}}\;S_n \ge 0\\
\frac{-\lvert f_j\rvert\sqrt P\theta_{j,n}S_n}{\sum\nolimits_{j \in {\cal J}_{o,n}}{\lvert f_j \rvert^2\theta_{j,n}^2}}>0, &\text{otherwise}
\end{cases}
\end{equation}
for each $j \in {\cal J}_{o,n}$, which implies that each terrestrial interference portion assigned to available BS $n$ is ensured to be suppressed, and can be completely cancelled if $S_n \ge 0$.

\begin{remark}
For the proposed distributed CB design, if the cooperation size $M$ is small, it may occur that $\Omega_j = \emptyset$ for some occupied BS $j$'s and/or ${\cal J}_{o,n}= \emptyset$ for some available BS $n$'s. In the former case, such occupied BSs do not need to split their interference to the UAV due to the absence of available BSs in their first $M$-tier neighborhood. In the latter case, such available BSs cannot perform the ITC and should simply set $v_{n,u}=\sqrt{P}$ and $v_{n,j}=0, j \in {\cal J}_o$, as in the conventional CB without ITC. For convenience, we assume in the rest of this section that $\Omega_j \ne \emptyset, \forall j \in {\cal J}_o$ and ${\cal J}_{o,n} \ne \emptyset, \forall n \in \Omega$, i.e., there exists at least one available (occupied) BS in the first $M$-tier neighborhood of an occupied (available) BS.
\end{remark}

\subsection{Interference Splitting}\label{IntSplit}
In this subsection, we address how to determine the interference splitting ratios $\{\theta_{j,n}\}_{n \in \Omega_j}$ at each occupied BS $j$. Depending on the practical requirement on the BSs' cooperation complexity and delay, we consider the following two protocols, namely open-loop and closed-loop, catered to delay-sensitive and delay-tolerant UAV communications, respectively.

\textbf{Open-loop protocol}: For delay-sensitive UAV communications (e.g., flying UAV with high speed), we consider an open-loop protocol, where each occupied BS only sends to their corresponding available BSs in $\Omega_j$ the splitting information once. In this case, each occupied BS has no prior knowledge about the CSI of the available BSs with the UAV. As such, each occupied BS $j \in {\cal J}_o$ equally splits its interference to the UAV into $\lvert \Omega_j \rvert$ portions, i.e., $\theta_{j,n} = 1/\lvert \Omega_j \rvert, n \in \Omega_j$.

\textbf{Closed-loop protocol}: In the previous open-loop protocol, the terrestrial interference is split regardless of the ITC capabilities of different available BSs, i.e., $S_n, n \in \Omega$. As a result, an available BS with low ITC capability may be heavily loaded with large terrestrial interference portions; while that with high ITC capability may be under-exploited with small terrestrial interference portions assigned. Fortunately, in delay-tolerant UAV communications (e.g., UAV hovering at a fixed location), the occupied BSs and their corresponding available BSs may be allowed to exchange information multiple times for refining the interference splitting ratios. Motivated by this, we propose a closed-loop protocol as follows. According to (\ref{res.int}), each available BS $n \in \Omega$ is capable of cancelling all terrestrial interference portions assigned to it if $S_n \ge 0$. As such, an iterative load balancing algorithm is designed to assign larger (smaller) interference splitting ratios to the available BSs with higher (lower) values of $S_n$, until a maximum number of iterations is reached.

Specifically, denote by $\theta_{j,n}^l \ge 0$ the interference splitting ratio in the $l$-th round of information exchange between occupied BS $j$ and available BS $n$ (in $\Omega_j$). Then, the ITC capability of available BS $n$ in this round is given by $S_n^l \triangleq {\lvert f_n \rvert}^2-\sum\nolimits_{j \in {\cal J}_{o,n}} {{\lvert f_j\theta_{j,n}^l \rvert}^2}$. In our proposed protocol, each available BS $n \in \Omega$ computes its current ITC capability $S_n^l$ and equally splits it among all occupied BSs in ${\cal J}_{o,n}$, each denoted as $A_n^l = S_n^l/\lvert {\cal J}_{o,n} \rvert$, then broadcasts $A_n^l$ to them for updating their respective interference splitting ratios. Obviously, if $A_n^l \ge 0$, each occupied BS $j \in {\cal J}_{o,n}$ can further assign $A_n^l$ to available BS $n$ for ITC, without changing the positive sign of $S_n^l$. Define $\Omega^{{\text{s}},l}_j \triangleq \left\{n\left| n \in \Omega_j, A_n^l > 0\right.\right\}$ as the set of available BSs in $\Omega_j$ with surplus ITC capabilities in the $l$-th round, and $\Omega^{{\text{d}},l}_j \triangleq \Omega_j \backslash \Omega^{{\text{s}},l}_j$ as the set of available BSs in $\Omega_j$ with deficit ITC capabilities in the $l$-th round. If $\Omega^{{\text{s}},l}_j \ne \emptyset$ and $\Omega^{{\text{d}},l}_j \ne \emptyset$, i.e., both ITC-surplus and ITC-deficit BSs exist in $\Omega_j$, we can increase $\theta_{j,n}^l, n \in \Omega^{{\text{s}},l}_j$ and  at the same time decrease $\theta_{j,n}^l, n \in \Omega^{{\text{d}},l}_j$ to better balance their ITC loads, i.e., $\theta_{j,n}^{l+1} \ge \theta_{j,n}^l, n \in \Omega^{{\text{s}},l}_j$ and $\theta_{j,n}^{l+1} \le \theta_{j,n}^l, n \in \Omega^{{\text{d}},l}_j$. Otherwise, if $\Omega^{{\text{s}},l}_j = \emptyset$ or $\Omega^{{\text{d}},l}_j = \emptyset$, each occupied BS $j$ does not update its interference splitting ratios, i.e., $\theta_{j,n}^{l+1} = \theta_{j,n}^{l}, n \in \Omega_j$.

For each occupied BS $j$ with $\Omega^{{\text{s}},l}_j \ne \emptyset$ and $\Omega^{{\text{d}},l}_j \ne \emptyset$, the following constraints must be met in updating its interference splitting ratios, i.e.,
\begin{equation}\label{up.cond}
\lvert f_j \rvert^2(\theta_{j,n}^{l+1})^2 - \lvert f_j \rvert^2(\theta_{j,n}^{l})^2 \le A_n^l, \forall n \in \Omega^{{\text{s}},l}_j,
\end{equation}
in order to preserve the positive signs of $S_n^l, n \in \Omega^{{\text{s}},l}_j$ in the next round. By some manipulations, (\ref{up.cond}) can be shown to be equivalent to
\begin{equation}\label{up.rule}
\theta _{j,n}^{l+1} - \theta _{j,n}^l \le \sqrt {\frac{A_n^l}{{\lvert f_j \rvert}^2} + (\theta _{j,n}^l)^2}-\theta _{j,n}^l\triangleq \Delta\theta_{j,n}^l, \forall n \in \Omega^{{\text{s}},l}_j.
\end{equation}

Based on (\ref{up.rule}), we define $Q_j^l = \sum\nolimits_{k \in \Omega^{{\text{s}},l}_j}{\Delta\theta_{j,k}^l}$, which denotes the maximum sum interference splitting ratio that the ITC-surplus BSs in $\Omega_j$ can further accommodate. For convenience, we refer to $Q_j^l$ as the maximum ITC quota for the ITC-surplus BSs in $\Omega_j$ in the $l$-th round. Apparently, the total decrease in $\theta_{j,n}^l, n \in \Omega^{{\text{d}},l}_j$ cannot exceed the maximum quota $Q_j^l$. To ensure this condition, the ratios $\theta_{j,n}^l, n \in \Omega^{{\text{d}},l}_j$ can be updated in an alternating manner, which follows the ascending order of $A_n^l, n \in \Omega^{{\text{d}},l}_j$, i.e., an available BS in $\Omega^{{\text{d}},l}_j$ with smaller $A_n^l$ has a higher priority to consume the quota. Specifically, each $\theta_{j,n}^l, n \in \Omega^{{\text{d}},l}_j$ is reduced to
\begin{equation}\label{new.ratios1}
\theta_{j,n}^{l+1} = \mathop {\max} \left\{0,\theta_{j,n}^l-Q_j^l\right\},
\end{equation}
and the maximum ITC quota is updated as $Q_j^l - (\theta_{j,n}^l-\theta_{j,n}^{l+1})$ accordingly. This update proceeds until $Q_j^l=0$ or all available BSs in $\Omega^{{\text{d}},l}_j$ have been updated.

Define $R_j^l \triangleq \sum\nolimits_{n \in \Omega^{{\text{d}},l}_j}\theta_{j,n}^{l}-\sum\nolimits_{n \in \Omega^{{\text{d}},l}_j}\theta_{j,n}^{l+1} \le Q_j^l$, i.e., the total decrease in the interference splitting ratios for the ITC-deficit BSs in $\Omega^{{\text{d}},l}_j$ after the above update. Since $\sum\nolimits_{n \in \Omega^{{\text{d}},l}_j}\theta_{j,n}^{l} + \sum\nolimits_{n \in \Omega^{{\text{s}},l}_j}\theta_{j,n}^l = \sum\nolimits_{n \in \Omega^{{\text{d}},l}_j}\theta_{j,n}^{l+1} + \sum\nolimits_{n \in \Omega^{{\text{s}},l}_j}\theta_{j,n}^{l+1}=1$, we have $\sum\nolimits_{n \in \Omega^{{\text{s}},l}_j}\theta_{j,n}^{l+1}-\sum\nolimits_{n \in \Omega^{{\text{s}},l}_j}\theta_{j,n}^l=R_j^l$, i.e., $R_j^l$ is the actual ITC quota for the ITC-surplus BSs in $\Omega_j$ in the $l$-th round. Similarly, we can also determine $\theta_{j,n}^{l+1}, n \in \Omega^{{\text{s}},l}_j$ in an alternating manner, which, however, follows the descending order of $A_n^l, n \in \Omega^{{\text{s}},l}_j$ to maximally exploit the available BSs with higher ITC capabilities. Specifically, each $\theta_{j,n}^l, n \in \Omega^{{\text{s}},l}_j$ is increased to
\begin{equation}\label{new.ratios2}
\theta_{j,n}^{l+1} = \mathop {\min} \left\{1, \theta_{j,n}^l + \mathop {\min} \left\{\Delta\theta_{j,n}^l, R_j^l\right\}\right\},
\end{equation}
and the actual ITC quota is updated as $R_j^l - (\theta_{j,n}^{l+1}-\theta_{j,n}^l)$ accordingly. This update proceeds until $R_j^l=0$.

After computing $\theta_{j,n}^{l+1}, n \in \Omega_j$, each occupied BS $j$ broadcasts the updated channel coefficients $f_j\theta_{j,n}^{l+1}, n \in \Omega_j$ to the available BSs in $\Omega_j$, and each available BS $n \in \Omega_j$ updates its ITC capability based on $S_n^{l+1} \triangleq {\lvert f_n \rvert}^2-\sum\nolimits_{j \in {\cal J}_{o,n}} {{\lvert f_j\theta_{j,n}^{l+1} \rvert}^2}$. The information exchange proceeds until a maximum round of exchange is reached. The initial interference splitting ratios, denoted by $\{\theta_{j,n}^1\}$, can be set according to the open-loop protocol, i.e., $\theta_{j,n}^1 = 1/\lvert \Omega_j \rvert, n \in \Omega_j, j \in {\cal J}_o$. The proposed closed-loop protocol ensures that the ITC-deficit BSs are assigned with non-increasing terrestrial interference portions, while preserving the signs of the capabilities of the ITC-surplus BSs, thus helping to improve the ITC loads of all available BSs. Moreover, we can obtain the following proposition.
\begin{proposition}\label{conv}
With the proposed closed-loop protocol and ITC, the total amount of terrestrial interference at the UAV receiver is non-increasing over $l$.
\end{proposition}
\begin{IEEEproof}
Consider two consecutive rounds $l$ and $l+1$. According to (\ref{res.int}), for any arbitrary occupied BS $j \in {\cal J}_o$, the amplitude of its residual interference to the UAV due to ITC in the $l$-th round, denoted as $\alpha_j^l$, is given by
\begin{equation}
\begin{split}
\alpha_j^l &= \sum\limits_{n \in \Omega_j^{{\text{d}},l}}{\frac{- \lvert f_j \rvert\sqrt P \theta_{j,n}^lS_n^l}{\sum\limits_{j \in {\cal J}_{o,n}}{{\lvert f_j \rvert}^2}\theta_{j,n}^{l2}}}\\
&= \sum\limits_{n \in \Omega_j^{{\text{d}},l}}{\lvert f_j \rvert \sqrt P \theta _{j,n}^l \cdot \left(1-\frac{{\lvert f_n \rvert}^2}{\sum\limits_{j \in {\cal J}_{o,n}} {{\lvert f_j \rvert}^2\theta_{j,n}^{l2}}}\right)},
\end{split}
\end{equation}
where the second equality is due to $S_n^l = {\lvert f_n \rvert}^2-\sum\nolimits_{j \in {\cal J}_{o,n}} {{\lvert f_j\theta_{j,n}^l \rvert}^2}$. Similarly, we can obtain
\begin{equation}
\alpha_j^{l+1} = \sum\limits_{n \in \Omega_j^{{\text{d}},l+1}}{\lvert f_j \rvert \sqrt P \theta _{j,n}^{l+1} \cdot \left(1-\frac{{\lvert f_n \rvert}^2}{\sum\limits_{j \in {\cal J}_{o,n}} {{\lvert f_j \rvert}^2\theta_{j,n}^{(l+1)2}}}\right)}.
\end{equation}
Moreover, the proposed distributed CB design ensures $\theta_{j,n}^l \ge \theta_{j,n}^{l+1}, \forall n \in \Omega_j^{{\text{d}},l}$, and thus $\Omega_j^{{\text{d}},l+1} \subseteq \Omega_j^{{\text{d}},l}$ (because an ITC-deficit BS in the $l$-th round may become ITC-surplus in the $(l+1)$-th round thanks to the reduced ITC load assigned to it). Then, it follows from $\Omega_j^{{\text{d}},l+1} \subseteq \Omega_j^{{\text{d}},l}$ that $\theta_{j,n}^l \ge \theta_{j,n}^{l+1}, \forall n \in \Omega_j^{{\text{d}},l+1}$, which gives rise to
\begin{equation}\label{conv.ineq}
1-\frac{{\lvert f_n \rvert}^2}{\sum\limits_{j \in {\cal J}_{o,n}} {{\lvert f_j \rvert}^2\theta_{j,n}^{l2}}} \ge 1-\frac{{\lvert f_n \rvert}^2}{\sum\limits_{j \in {\cal J}_{o,n}} {{\lvert f_j \rvert}^2\theta_{j,n}^{(l+1)2}}}, \forall n \in \Omega_j^{{\text{d}},l+1}.
\end{equation}
Therefore, it must hold that
\[\alpha_j^l \ge \sum\limits_{n \in \Omega_j^{{\text{d}},l+1}}{\lvert f_j \rvert \sqrt P \theta _{j,n}^l \cdot \left(1-\frac{{\lvert f_n \rvert}^2}{\sum\nolimits_{j \in {\cal J}_{o,n}} {{\lvert f_j \rvert}^2\theta_{j,n}^{l2}}}\right)} \ge \alpha_j^{l+1},\]
where the first inequality is due to $\Omega_j^{{\text{d}},l+1} \subseteq \Omega_j^{{\text{d}},l}$, while the second inequality is due to $\theta_{j,n}^l \ge \theta_{j,n}^{l+1}, \forall n \in \Omega_j^{{\text{d}},l+1}$ and (\ref{conv.ineq}). The proof is thus completed.
\end{IEEEproof}

\subsection{Algorithm Implementation}
Next, we discuss how to practically implement the proposed distributed CB design. To initiate the protocol, the UAV first broadcasts a beacon signal over its assigned RB for all available and occupied BSs in the interfering BS region $D_u$ to measure their downlink channel coefficients with the UAV, i.e., $f_j, j \in {\cal J}_o \cup \Omega$. Here, it is assumed that the cellular network adopts the time division duplex (TDD) mode, such that the channel reciprocity holds in practice. Suppose that each occupied BS is aware of the list of all available BSs in its $M$-tier neighborhood, which is enabled by the X2 interface in LTE\cite{3GPP36423}. Then, each occupied BS $j \in {\cal J}_o$ broadcasts its initial channel coefficients $f_j\theta_{j,n}^1, n \in \Omega_j$ to the available BSs in $\Omega_j$ to start the information exchange. Assume that the maximum allowable round of information exchange is $L$. The open-loop protocol should be implemented only with $L=1$. After $L$ rounds of information exchange, each occupied BS $j \in {\cal J}_o$ sends the data symbol of its served terrestrial UE (i.e., $x_j$) to the available BSs in $\Omega_j$ with positive interference splitting ratios. The whole process of the distributed CB protocol is summarized in Algorithm 1.

In general, the proposed distributed CB design cannot achieve the same performance as its centralized counterpart. The fundamental reason lies in that there is no global cooperation among all available BSs; thus, the distributed design cannot provide full CB gains as in the centralized one. In addition, under a limited cooperation size of $M$, the interference from an occupied BS can only be cancelled by a subset of available BSs in $D_u$ that are within its neighborhood, as compared to all available BSs in the centralized design. Nonetheless, the proposed distributed CB design can still effectively suppress the terrestrial interference and outperform the conventional CB without applying ITC, as will be shown in the next section.
\begin{algorithm}
  \caption{Distributed CB protocol}\label{Alg1}
  \begin{algorithmic}[1]
    \State The UAV sends a beacon signal to inform the BSs in $D_u$ its existence over its assigned RB.
    \State Each occupied/available BS measures its downlink channel coefficient with the UAV based on the received signal.
    \State Initiate $l=1$ and $\theta_{j,n}^l=1/\lvert \Omega_j \rvert, n \in \Omega_j, j \in {\cal J}_o$.
    \State Each occupied BS $j \in {\cal J}_o$ broadcasts the initial channel coefficients $f_{j,n}^l=f_j\theta_{j,n}^l, n \in \Omega_j$ to all available BSs in $\Omega_j$.
    \While {$l < L$}
    \State Each available BS $n \in \Omega$ computes $A_n^l$ and broadcasts it to the occupied BSs in ${\cal J}_{o,n}$.
    \State Each occupied BS $j \in {\cal J}_o$ updates the interference splitting ratios as $\theta_{j,n}^{l+1}, n \in \Omega_j$ as in Section \ref{IntSplit}, and then broadcasts $f_{j,n}^{l+1}=f_j\theta_{j,n}^{l+1}, n \in \Omega_j$ to all available BSs in $\Omega_j$.
    \State Update $l=l+1$.
    \EndWhile
    \State Each occupied BS $j \in {\cal J}_o$ shares its served terrestrial UE's data symbol $x_j$ with the available BSs in $\Omega_j$ with non-zero $\theta_{j,n}^L$.
    \State Each available BS $n \in \Omega$ computes the phases and amplitude of its beamforming weights based on Proposition \ref{opt.phase} (by replacing ${\cal J}_o$ with ${\cal J}_{o,n}$) and (\ref{solution5})-(\ref{solution6}) (by replacing $\theta_{j,n}$ with $\theta_{j,n}^L$), respectively, and then initiates the downlink transmission to the UAV.
  \end{algorithmic}\vspace{-3pt}
\end{algorithm}

\section{Simulation Results}\label{sim}
In this section, simulation results are provided to evaluate the performance of our proposed downlink CB designs with ITC. Unless otherwise specified, the simulation settings are as follows. The tier of neighboring BSs is $q=1$ for the terrestrial ICIC. The RB assigned to the UAV consists of 12 consecutive subcarriers, with the subcarrier spacing being 15 kHz. The cell radius is $800$ m, and the height of BSs is set to be 25 m. The altitude of the UAV is fixed as 200 m. The carrier frequency $f_c$ is at $2$ GHz, and the noise power spectrum density at the receiver is $-164$ dBm/Hz including a 10 dB noise figure. The BS antenna pattern is assumed to be directional in the vertical plane but omnidirectional in the horizontal plane\footnote{Our proposed CB designs are also applicable to the case with sectorized antenna pattern of the BS in the horizontal plane.}. Specifically, we consider that the BS antenna pattern is synthesized by a uniform linear array (ULA) with 10 co-polarized dipole antenna elements\cite{ballanis2016antenna}. The antenna elements are placed vertically with half-wavelength spacing and electrically steered with 10-degree downtilt angle. The UAV-BS channels follow the probabilistic LoS/NLoS channel model based on the urban macro (UMa) scenario in \cite{3GPP36777} (see Tables B-1 and B-2 in \cite{3GPP36777} for the expressions of LoS probability and path-loss, respectively). We consider three tiers of cells centered at the cell underneath the UAV (named cell 1) to cover the interfering BS region $D_u$, and thus the total number of cells is $J=37$. The BS in cell 1 is assumed to be located at the origin without loss of generality. The UAV's horizontal location is fixed at (150 m, 420 m) in cell 1. The terrestrial UEs' locations are randomly generated in the $J$ cells, which can change the distribution and number of available BSs. All results shown in this section have been averaged over 200 random realizations of the terrestrial UEs' locations.

\begin{figure}[htb]
\centering
\includegraphics[width=3.2in]{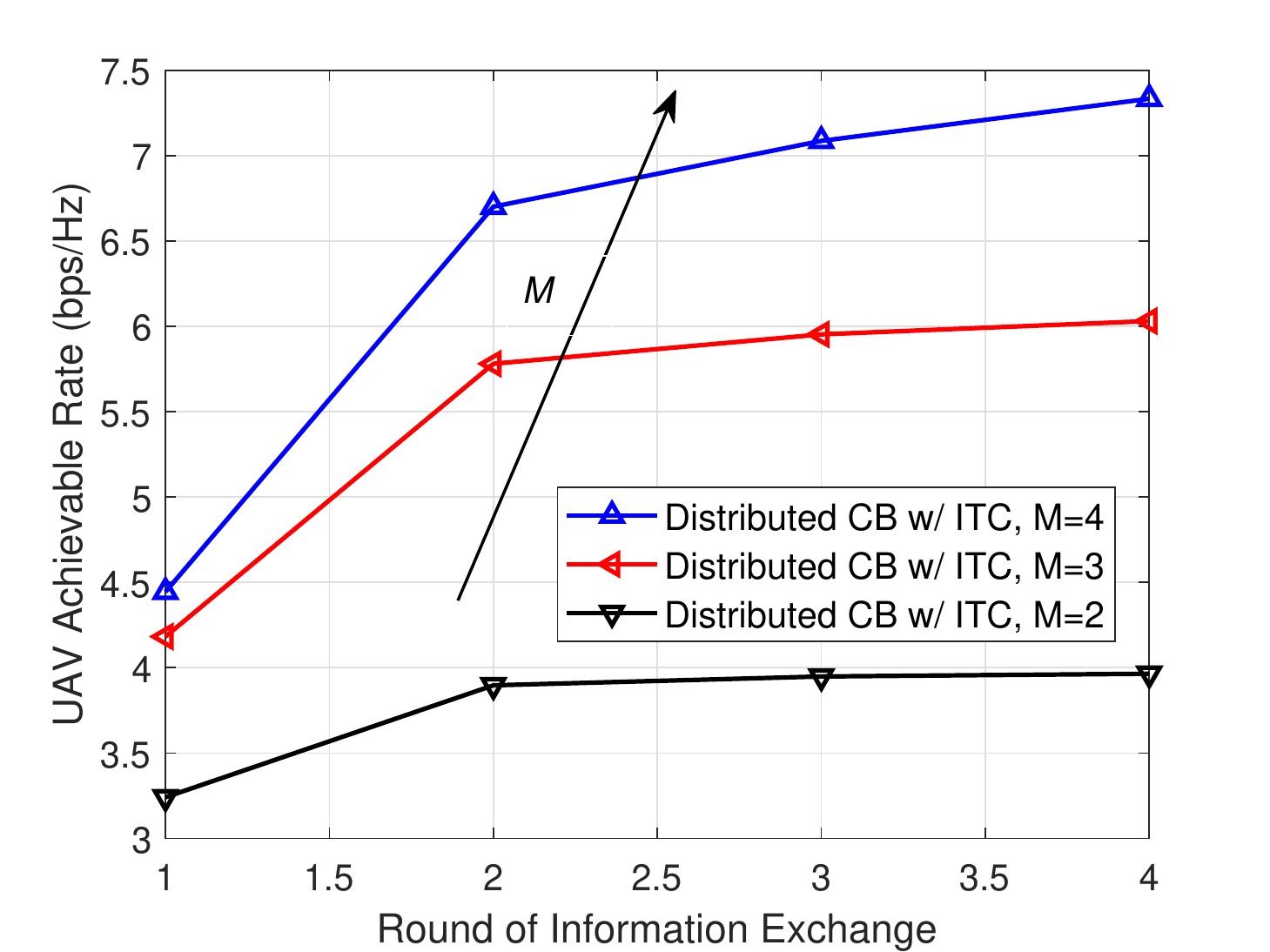}
\DeclareGraphicsExtensions.
\caption{UAV achievable rate versus round of information exchange.}\label{Loop_Thrpt_CB}
\end{figure}
First, Fig.\,\ref{Loop_Thrpt_CB} shows the UAV's achievable rate (defined as $\log_2(1+\gamma)$ in bits per second per Hertz (bps/Hz), where $\gamma$ denotes the UAV's achievable SINR in each scheme) by the proposed distributed CB with ITC versus the maximum information exchange round $L$, under different cooperation size of $M$. The BS transmit power and the number of terrestrial UEs are set to $P=30$ dBm and $K=7$, respectively. The UAV's achievable rate for $L=1$ corresponds to the open-loop protocol. It is observed from Fig.\,\ref{Loop_Thrpt_CB} that the UAV's achievable rate keeps increasing with $L$, thanks to the improved ITC performance, in accordance with Proposition \ref{conv}. Moreover, the UAV's achievable rate is observed to be improved by increasing the cooperation size $M$. This is expected as a larger cooperation size leads to a larger number of available BSs for ITC, and thus a stronger ITC capability.

\begin{figure}[htb]
\centering
\includegraphics[width=3.2in]{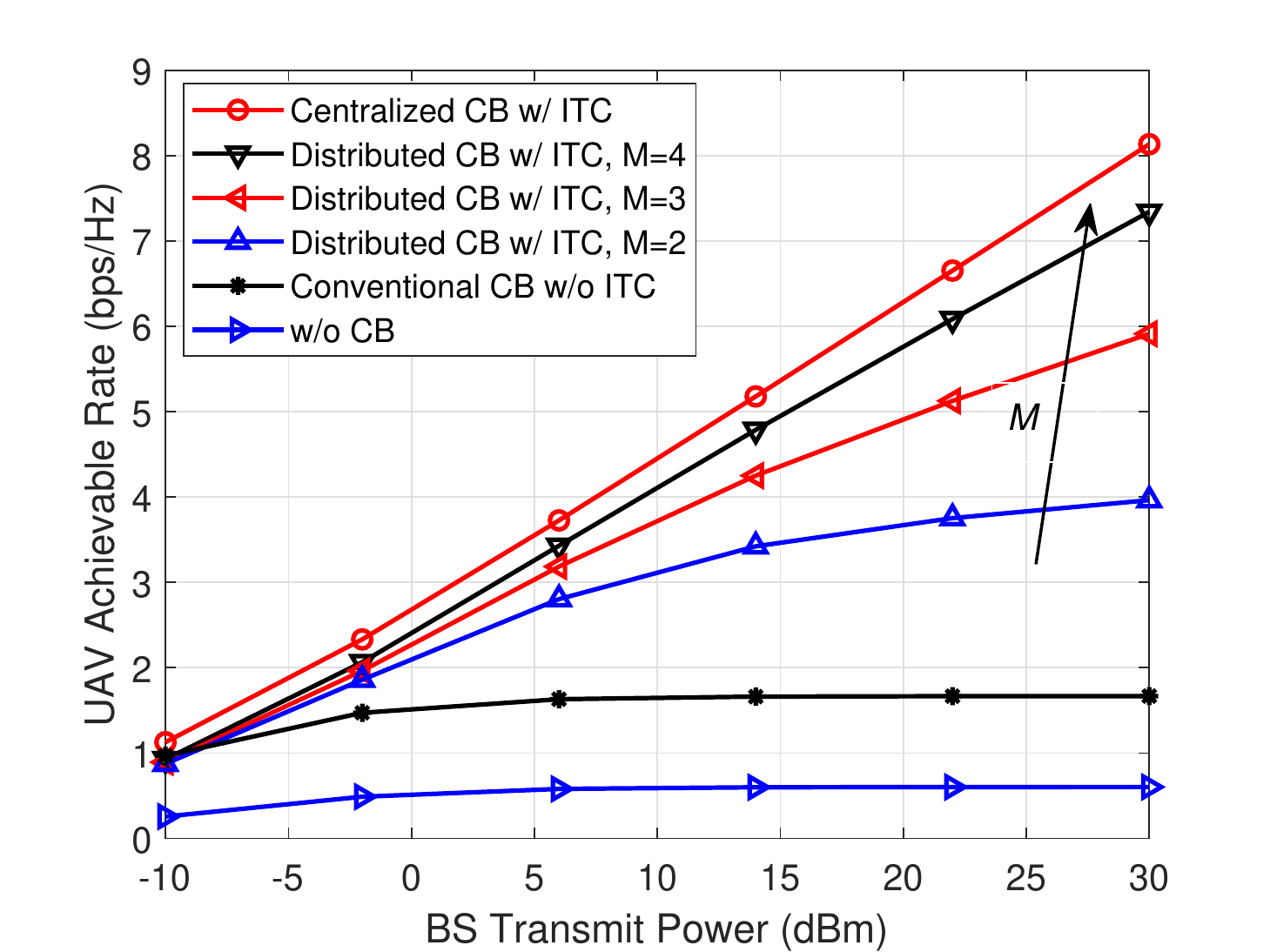}
\DeclareGraphicsExtensions.
\caption{UAV achievable rate versus BSs' transmit power.}\label{Pw_Thrpt_CB}
\end{figure}
Next, Fig.\,\ref{Pw_Thrpt_CB} shows the UAV's achievable rate versus the BS transmit power $P$ under different cooperation size of $M$. The maximum information exchange round is $L=3$. The total number of co-channel terrestrial UEs in $D_u$ is $K=7$. From Fig.\,\ref{Pw_Thrpt_CB}, it is observed that in the moderate-to-high transmit power regime, the proposed CB designs with ITC (centralized and distributed) significantly outperform the two benchmark schemes, namely, the scheme without CB and the conventional CB scheme without applying ITC, as described in Section \ref{new.uav}. In particular, for the two benchmark schemes, increasing $P$ can only bring marginal improvement of the UAV's achievable rate. The reason is that the terrestrial interference power increases with $P$ at a rate comparable to the UAV's received signal power in the high transmit power regime. Hence, the UAV's achievable rate is severely limited by the terrestrial interference. In contrast, for the proposed CB scheme with ITC, increasing $P$ leads to dramatic improvement of the UAV's achievable rate. This is because with ITC, the terrestrial interference can be eliminated or substantially suppressed, and its residual power after ITC increases much more slowly with $P$ than the UAV's received signal power. However, in the low transmit power regime (e.g., $P=-10$ dBm), it is observed that all considered CB schemes yield comparable UAV's achievable rates. This is because the UAV's achievable rate is mainly affected by the receiver noise instead of the terrestrial interference when $P$ is small, thus diminishing the performance gain by ITC. Last, it is observed that with $M=4$, the performance gap between the centralized CB and the distributed CB is practically small over the whole range of transmit powers.

\begin{figure}[htb]
\centering
\includegraphics[width=3.2in]{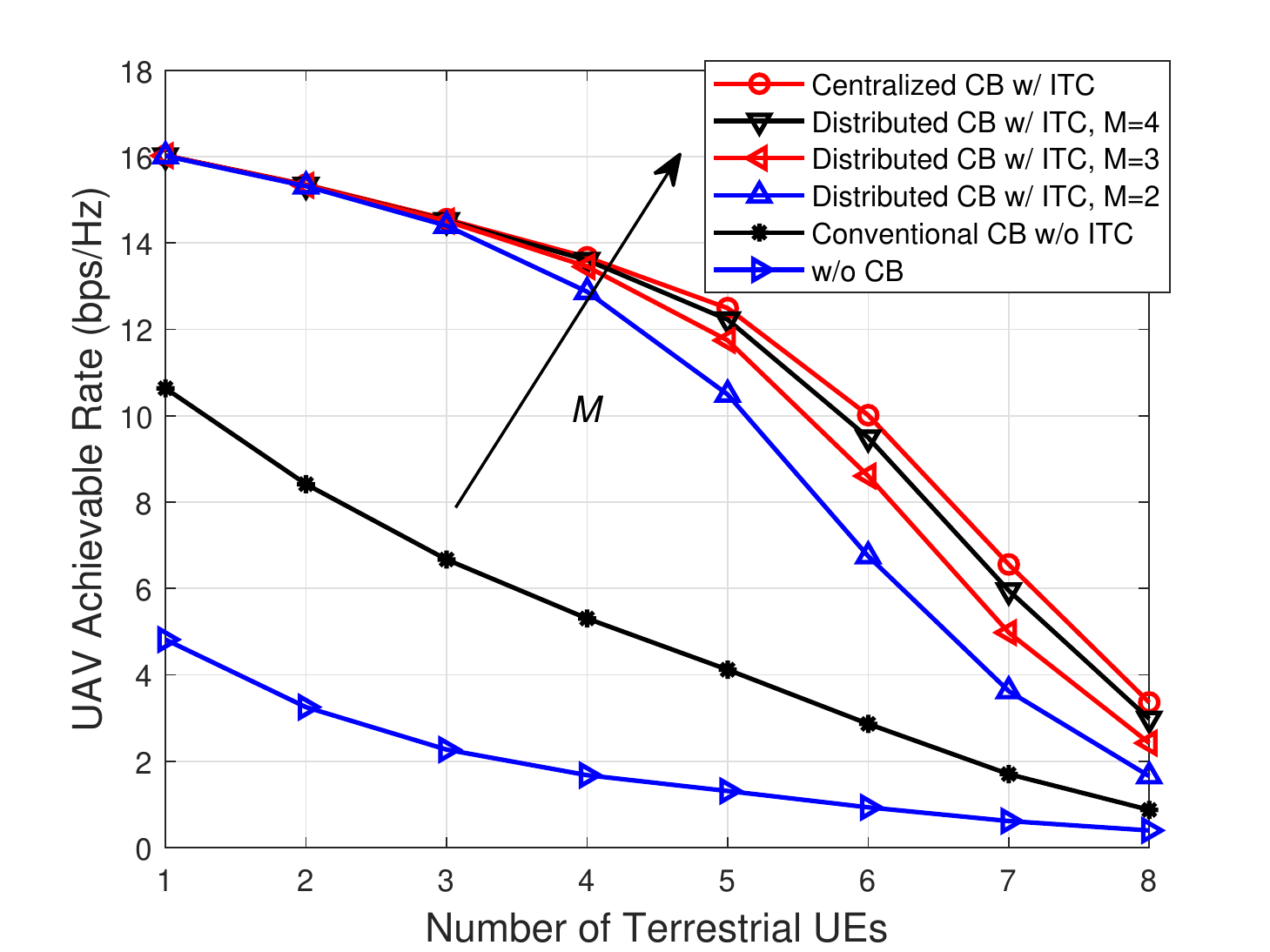}
\DeclareGraphicsExtensions.
\caption{UAV achievable rate versus number of terrestrial UEs.}\label{UE_Thrpt_CB}
\end{figure}
Fig.\,\ref{UE_Thrpt_CB} plots the UAV's achievable rate versus the number of co-channel terrestrial UEs (or occupied BSs) $K$ with $P=20$ dBm. It is observed that the UAV's achievable rates decrease with $K$ under all the considered schemes. On one hand, this is because the UAV suffers stronger aggregate terrestrial interference from more occupied BSs. On the other hand, this is due to the decreasing number of available BSs and hence the decreased CB gains for both the UAV's signal power enhancement and ITC. The decreased CB gain also accounts for the small performance gap between the two benchmark schemes under high terrestrial UE density. In contrast, the proposed CB designs with ITC are observed to provide significant performance gains over the two benchmark schemes even under high terrestrial UE density. Moreover, it is observed that the performance gap between the centralized CB and the distributed CB is still small. In particular, when the terrestrial UE density is low, the distributed CB can achieve almost the same performance as the centralized one even with a small cooperation size $M=2$. This implies that for small $K$, the terrestrial interference from an occupied BS only needs to be cancelled by its nearest available BSs, which further reduces the signaling overhead and delay.

\begin{figure}[htb]
\centering
\includegraphics[width=3.2in]{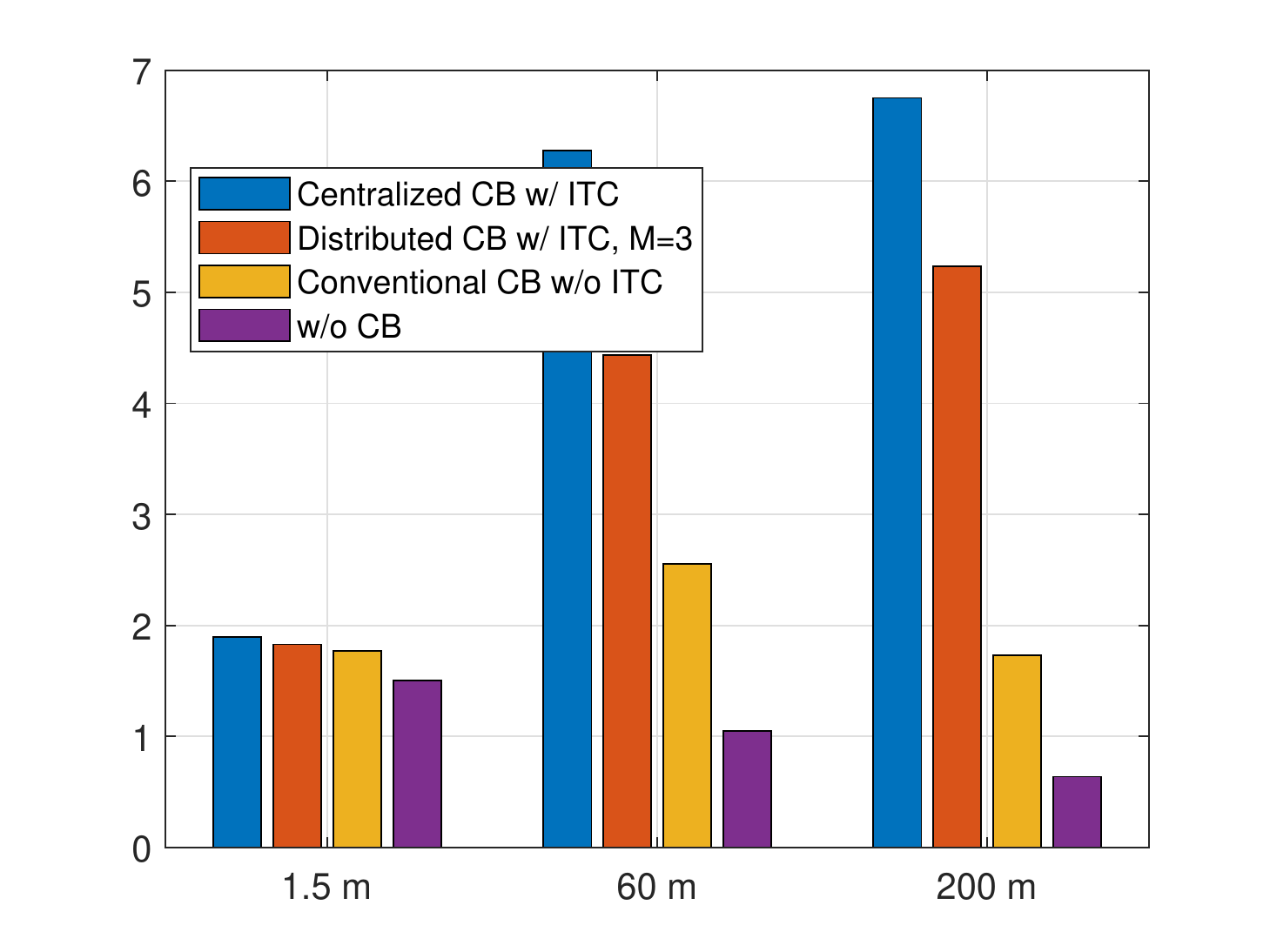}
\DeclareGraphicsExtensions.
\caption{UAV achievable rate versus UAV altitude.}\label{Ht_Thrpt_CB}
\end{figure}
Finally, we plot the UAV's achievable rates for the considered system with the UAV altitudes 1.5 m, 60 m, and 200 m in Fig.\,\ref{Ht_Thrpt_CB}. The BS transmit power and the number of terrestrial UEs are set to $P=20$ dBm and $K=7$, respectively. The BS cooperation size is $M=3$ in the distributed CB design. The case with UAV altitude 1.5 m may correspond to either a benchmark ground UE or a UAV in take-off/landing status. From Fig.\,\ref{Ht_Thrpt_CB}, it is observed that at the low altitude 1.5 m, the proposed CB designs with ITC yield only small performance gains over the conventional CB. This is expected as in this case, the UAV's received interference is much weaker than that at moderate or high altitude due to the similar terrestrial channel condition. However, with increasing UAV altitude, its channel condition with the ground BSs improves due to the decreased NLoS probability and path-loss exponent, thus enhancing both its received signal power and interference power. In particular, at the altitudes 60 m and 200 m, the UAV's achievable rate is observed to be worse than that at 1.5 m under the schemes without CB or with conventional CB. This implies that the UAV's achievable rate is mainly limited by the growth in the interference power at moderate-to-high altitude, even with the conventional CB applied. In contrast, the UAV's achievable rates by the proposed CB designs are observed to increase with the altitude, owing to the far less dominant interference power after ITC.

\section{Conclusion and Future Work}
This paper proposed a new CB scheme with ITC to mitigate the strong downlink interference to cellular-connected UAVs, by exploiting the cooperative transmission of co-channel interference for cancellation at the UAV receiver. The UAV's receive SINR was maximized via jointly optimizing the power allocations to balance between UAV signal power enhancement and terrestrial interference suppression, which is a new and fundamental trade-off revealed in the proposed CB design. We solved this problem optimally and also obtained useful insights into the above trade-off for maximizing the UAV's downlink SINR. To reduce the implementation complexity and overhead of the optimal CB design, we further proposed a distributed CB design, based on a novel divide-and-conquer approach requiring only local information exchange among BSs involved.
Simulation results showed that the proposed centralized and distributed CB designs with ITC significantly improve the UAV's downlink achievable rate over the conventional CB without applying ITC, especially when the BS transmit power is large or the terrestrial UE density is high. It was also shown that increasing the cooperation size helps enhance the performance of the distributed CB design, at the cost of more complexity and processing delay. This paper can be extended in several promising directions for future work. For example, more sophisticated CB design is needed in the absence of available BSs, e.g., when the terrestrial UE density is extremely high. In addition, it is also interesting to consider the more general case with multiple co-channel UAVs and/or multi-antenna UAVs, where the trade-offs between UAV signal enhancement and interference cancellation are more intricate to be characterized for the optimal CB design.

\appendix[Proof of Proposition \ref{single.opt}]
First, by introducing a slack variable $\eta$, (P3) can be equivalently reformulated as the following problem:
\begin{subequations}\label{opA1}
\begin{align}
\nonumber \mathop {\max}\limits_{\eta,v_j,v_u}&\; \eta \\
\text{s.t.}\;\;&\lvert f_a \rvert^2v^2_u \ge \eta(\sigma ^2 + (\lvert f_o \rvert\sqrt{P} -\lvert f_a \rvert v_j)^2),\label{opA1a}\\
&v_j^2  + v_u^2 \le P.\label{opA1b}
\end{align}
\end{subequations}

The Lagrangian of problem (\ref{opA1}) is given by
\begin{equation}
\begin{split}
{\cal L}(\eta,v_j,v_u,\lambda,\nu)=&\eta+\lambda(\lvert f_a \rvert^2v^2_u \!-\! \eta\sigma ^2 \!-\! \eta(\lvert f_o \rvert\sqrt{P}\!-\!\lvert f_a \rvert v_j)^2)\\
&+\nu(P-v_j^2-v_u^2),\nonumber
\end{split}
\end{equation}
where $\lambda \ge 0$ and $\nu \ge 0$ are the dual variables associated with constraints (\ref{opA1a}) and (\ref{opA1b}) in problem (\ref{opA1}), respectively.

By taking the derivative of the Lagrangian of problem (\ref{opA1}) and setting it to zero, we can obtain the following KKT conditions, i.e.,
\begin{align}
\frac{\partial L}{\partial{v_j}} &= 2\lambda \eta \lvert f_a \rvert (\lvert f_o \rvert\sqrt P  - \lvert f_a \rvert {v_j}) - 2\nu v_j = 0,\label{kkt1}\\
\frac{\partial L}{\partial{v_u}} &= 2\lambda\lvert f_a \rvert^2v_u - 2\nu v_u = 0,\label{kkt2}\\
\frac{\partial L}{\partial{\eta}} &= 1-\lambda(\lvert f_o \rvert\sqrt{P} -\lvert f_a \rvert v_j)^2-\lambda\sigma^2 = 0.\label{kkt3}
\end{align}

As $v_u > 0$, the KKT condition (\ref{kkt2}) can be simplified as $\nu=\lambda\lvert f_a \rvert^2$. By substituting this into (\ref{kkt1}), we have $\lambda \eta \lvert f_a \rvert (\lvert f_o \rvert\sqrt P  - \lvert f_a \rvert {v_j})=\lambda\lvert f_a \rvert^2v_j$. It follows from (\ref{kkt3}) that $\lambda>0$; thus, we can obtain
\begin{equation}\label{eqA1}
v_j = \frac{\eta\lvert f_o \rvert \sqrt P}{(\eta+1)\lvert f_a \rvert}.
\end{equation}

Moreover, it is easy to verify that the equality in (\ref{opA1b}) must hold at the optimality of problem (\ref{opA1}), since otherwise we can increase $v_u$ to achieve a larger objective value. As a result, we have $v_u^2 = P - v_j^2$. Similarly, the equality in (\ref{opA1b}) also holds at the optimality of problem (\ref{opA1}), i.e., $\lvert f_a \rvert^2v^2_u = \eta(\sigma ^2 + (\lvert f_o \rvert\sqrt{P} -\lvert f_a \rvert v_j)^2)$. By plugging $v_u^2 = P - v_j^2$ into this equality, we arrive at
\begin{equation}\label{eqA2}
\lvert f_a \rvert^2(P - v_j^2) = \eta(\sigma ^2 + (\lvert f_o \rvert\sqrt{P} -\lvert f_a \rvert v_j)^2).
\end{equation}

Next, by plugging (\ref{eqA1}) into (\ref{eqA2}) and after some manipulations, we can obtain the following quadratic equation of $\eta$, i.e.,
\begin{equation}\label{eqA3}
\sigma^2\eta^2 + (\sigma^2+P\lvert f_o \rvert^2-P\lvert f_a \rvert^2)\eta-\lvert f_a \rvert^2P=0.
\end{equation}
Notice that the equation in (\ref{eqA3}) has one negative root and one positive root. Evidently, $\eta^\star$ should be the unique positive root of (\ref{eqA3}), as given in (\ref{opt.eta}). By substituting (\ref{opt.eta}) into (\ref{eqA1}), we can obtain (\ref{solution1}). Thus, the proof is completed.

\bibliography{UAV_CB}
\bibliographystyle{IEEEtran}

\end{document}